\documentclass[%
 reprint,
 onecolumn,
superscriptaddress, 
showkeys,
 amsmath,amssymb,
 aps,
]{revtex4}
\usepackage{graphicx,subcaption} 
\usepackage{subcaption}
\usepackage{ragged2e}
\DeclareCaptionJustification{justified}{\justifying}
\usepackage{dcolumn}
\usepackage{bm}
\usepackage{orcidlink}
\hypersetup{hidelinks}
\hypersetup{
    colorlinks = true,
    citecolor = blue,
    linkcolor = blue,
     urlcolor = blue,
}

\newcommand{\RNum}[1]{\uppercase\expandafter{\romannumeral #1\relax}}

\begin{document}

\preprint{APS/123-QED}
\title{Interplay of the Fano-Andreev Effect and Nonlinear Screening in Electric and Thermoelectric Transport through a T-Shaped Double Quantum Dot}%
\author{Sachin Verma\,\orcidlink{0000-0001-9490-8404}}
\altaffiliation[Email: ]{sachin@manit.ac.in}
\affiliation{Department of Physics\text{,} Maulana Azad National Institute of Technology Bhopal, 462003, Madhya Pradesh, India.}



\date{11 November, 2026}%
\begin{abstract}
We present a theoretical study of weakly nonlinear electric and thermoelectric transport through a T-shaped double quantum dot coupled to normal and superconducting leads. Nonlinear screening is treated within the characteristic-potential formalism, in which an internal electrostatic potential accounts for nonequilibrium charge redistribution on the lead-coupled central dot. The side-coupled dot induces the Fano-Andreev effect, which, together with nonlinear screening, significantly modifies the transport properties. Under voltage bias, the screening-induced level renormalization produces a pronounced bias asymmetry in the Andreev differential conductance. In response to a pure thermal bias, increasing the interdot coupling weakens the level renormalization and suppresses the quasiparticle thermoelectric response, leading to reduced thermovoltage and nonlinear Seebeck coefficient. When voltage and thermal biases are applied simultaneously, a finite subgap cross thermoelectric response emerges beyond the linear regime. The sign of this response is governed by the competition between the energy-dependent transport contribution, modified by the Fano-Andreev effect, and the nonlinear screening contribution, thereby determining whether a thermal bias enhances or suppresses the low-bias Andreev current.
\end{abstract}
\maketitle
\section{Introduction}
\label{sec:1}
The study of thermoelectric and heat transport in hybrid superconducting mesoscopic (low-dimensional) systems, including quantum dots (QDs), nanowires, and molecular junctions, has gained significant attention with the development of quantum thermodynamics and nanoscale thermal management~\cite{Giazotto2009,Benenti2016,Sanchez2016,Pekola2021,Arrachea2025}. Early research on thermoelectric effects was primarily motivated by the goal of enhancing energy harvesting and improving conversion efficiency~\cite{Heremans2013,Whitney2014,Cangemi2024}. In this context, hybrid superconducting  quantum dot systems offer efficient and promising routes toward nanoscale heat engines, coolers, thermal diodes, sensors, and other emerging superconducting quantum technologies operating at millikelvin temperatures~\cite{Giazotto2009,Benenti2016,Manzano2020}. These advances in mesoscopic caloritronics have further demonstrated that thermoelectric phenomena at the nanoscale are not only of practical relevance but also provide deep insight into fundamental quantum processes. In phase-coherent conductors, the thermoelectric response serves as a sensitive probe of particle-hole asymmetry, electronic correlations, and the underlying spectral properties of quantum states. Hybrid structures incorporating Bardeen-Cooper-Schrieffer (BCS) or s-wave superconductors are particularly attractive in this regard. Such systems enable the generation and control of entangled electron pairs via Cooper-pair splitting~\cite{Cao2015,Sanchez2018,Hussein2019,Arora2025} and provide a thermoelectric probe for topological~\cite{Leijnse2014,Lopez2014,Trocha2024,Trocha2025,Smirnov2015}, and collective excitations~\cite{Kamp2021}. For a comprehensive and recent review covering fundamental aspects of thermoelectric and heat transport in hybrid superconducting mesoscopic systems, see Ref.~\cite{Arrachea2025}.\\
Superconductor-quantum dot hybrid mesoscopic devices provide a highly tunable platform in which quantum confinement, strong Coulomb interaction, and proximity-induced superconducting correlations coexist~\cite{Meden2019,Martin2011}. For example, when a quantum dot is coupled to a ferromagnetic or normal metallic lead on one side and to a superconducting lead on the other (N/F-QD-S junctions), Andreev reflection induces coherent electron-hole mixing, leading to the formation of discrete Andreev bound states (ABSs) that govern subgap transport via Cooper pair tunneling; under a purely voltage bias, these processes dominate because single-particle or quasiparticle tunneling is suppressed within the superconducting energy gap~\cite{Martin2011}. For energies above the gap, transport is dominated by quasiparticle tunneling, where single-particle excitations contribute to charge and heat transport through the energy-dependent quasiparticle density of states~\cite{Krawiec2008,Hwang2015,Trocha2017}. The intrinsic particle-hole symmetry of Andreev reflection ensures a symmetric Andreev transmission probability, $T_A(\omega)=T_A(-\omega)$. 
Therefore, a purely thermal gradient cannot generate an Andreev thermoelectric response, since the contributions from electron-like and hole-like excitations cancel around the chemical potential~\cite{Hwang2016b,Verma2022,Arrachea2025}. Consequently, Andreev processes do not contribute to net heat and entropy transport, whereas quasiparticle tunneling can produce a finite thermoelectric response when particle-hole symmetry is broken, for example by tuning the quantum-dot energy level away from the Fermi energy~\cite{Hwang2015,Trocha2017}, Zeeman splitting~\cite{Ozaeta2014} or spin polarization of the ferromagnetic lead~\cite{Machon2013}. Under such conditions, quasiparticle transport gives rise to thermopower, enhanced by the strong energy dependence of the BCS density of states near the superconducting gap edges, while Andreev tunneling can influence thermoelectric properties indirectly under open-circuit conditions through the compensating current, modifying heat conductance, thermopower, and the corresponding figures of merit~\cite{Hwang2016b,Trocha2017,Verma2022}.
\begin{figure*}[!htb]
\centering
\includegraphics
[width=0.6\hsize]
{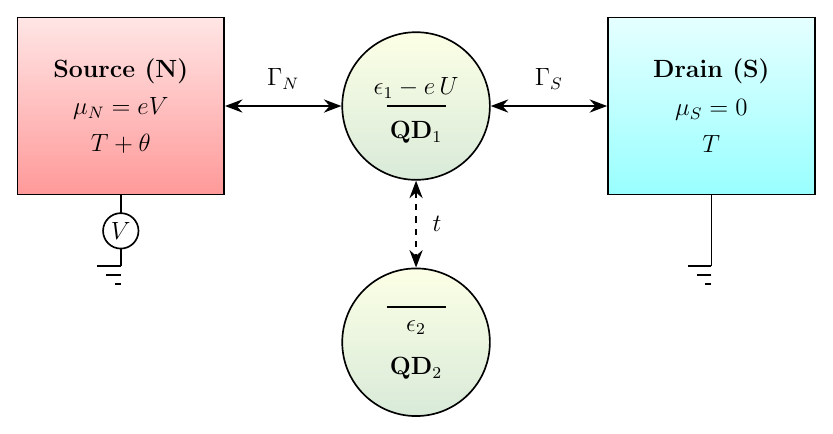}
\caption{Schematic of a T-shaped double quantum dot (DQD) hybrid junction coupled to a normal (N) and a superconducting (S) leads. The normal lead is biased by a voltage $V$, corresponding to a chemical potential $\mu_N=eV$, and held at a temperature $T+\theta$, where $\theta$ denotes the applied thermal bias relative to the background temperature $T$. The superconducting lead remains at equilibrium $(\mu_S=0,\;T)$. The lead-coupled central dot (QD$_1$) is tunnel-coupled to the leads $\alpha\in\{\mathrm{N},\mathrm{S}\}$ with coupling strengths $\Gamma_{\alpha}$ and to the side-coupled dot (QD$_2$) through the interdot coupling $t$, whereas QD$_2$ is not directly connected to either lead. Each dot contains a single energy level. The energy level of QD$_1$ ($\epsilon_1$) is renormalized by the bias-induced screening potential $U$, while the energy level of QD$_2$ ($\epsilon_2$) remains unchanged. Therefore, QD$_2$ acts as an energy-dependent scattering channel, giving rise to the Fano-Andreev effect.}
\label{fig:1}
\end{figure*}\\
More complex nanostructures have also been investigated to enhance thermoelectric performance. For example, coupled or triple quantum-dot systems connected to normal or ferromagnetic and superconducting leads exhibit quantum interference effects, such as Fano and Dicke resonances, as well as Coulomb interaction and Andreev reflection processes, which can significantly modify thermoelectric coefficients and improve the thermoelectric figure of merit~\cite{XU2016,Godzik2017,Yao2018a,Yao2018b,Yao2023}. Hybrid superconducting quantum-dot devices have also been proposed as nonlocal heat engines~\cite{Bhandari2018,Tabatabaei2020,Tabatabaei2022}.
In a related setup, hybrid normal-superconductor quantum-dot systems have been shown to operate as Andreev reflection engines, where coherent Cooper-pair processes enable quantum-enhanced power generation and efficiency~\cite{Manzano2023}. Multiterminal superconducting nanostructures with coupled quantum dots can further exhibit interference effects and molecular Andreev bound states that strongly influence both local and nonlocal thermoelectric transport ~\cite{Wysokinski2012,Michalek2016,Sonar2024,Sonar2025a,Gorsk2025}. In addition, Cooper pair splitter devices based on double quantum dots coupled to a superconductor and two normal leads can generate nonlocal thermoelectric currents driven by temperature gradients through crossed Andreev reflection~\cite{Cao2015,Sanchez2018,Kirsanov2019,Hussein2019,Tan2021,Jong2023,Arora2025}. Beyond conventional thermoelectric effects, additional phenomena such as spin and charge Seebeck diode behavior~\cite{Hwang2016a}, phase dependent heat transport~\cite{Kamp2019,Bauer2021,Hwang2024} and thermophase Seebeck effect~\cite{Kleeorin2016,Kumar2023,Gao2023,Kumar2025} have also been reported in hybrid superconducting-quantum dot devices. Hybrid superconductor–quantum dot nanostructures have also been proposed as efficient electronic coolers, where voltage- or thermal-driven transport-mediated by Andreev reflection and quasiparticle tunneling-enables refrigeration of normal metals, resonators, and electronic circuits with efficiencies approaching a significant fraction of the Carnot limit~\cite{Hwang2023,Verma2024,Tabatabaei2024,Cioni2025}. Pure electrical transport has also been studied in T-shaped DQD structures coupled to normal and superconducting leads. Early studies showed that the interplay of superconducting proximity, quantum interference, and electronic correlations strongly modifies Andreev transport, giving rise to characteristic Fano features in the subgap regime~\cite{Tanaka2008,Baranski2011,Calle2017}. Subsequent studies investigated the robustness of the Fano-Andreev effect against Coulomb correlations~\cite{Gonzalez2024} and the influence of superconductor-QD coupling on Fano interference~\cite{Calle2013}. Related studies also explored the effects of electron-boson interactions in T-shaped DQDs~\cite{Wu2006,Baranski2015,He2015}, as well as the influence
of Majorana modes on the Andreev conductance and Fano-type transport features~\cite{Majek2022,He2025}.\\
Despite significant progress, previous studies have largely focused on linear-response or purely electrical transport, without accounting for bias-induced internal electrostatic screening. Moreover, a temperature gradient, particularly when combined with a finite voltage bias, drives the system out of equilibrium, giving rise to nonlinear and cross-thermoelectric effects. Under simultaneous voltage and thermal biases, Andreev processes can contribute to a finite nonlinear thermoelectric response~\cite{Hwang2015,Hwang2017,Marchegiani2020}. In this work, we investigate weakly nonlinear electric and thermoelectric transport through a T-shaped double quantum dot (DQD) coupled to normal-metal and superconducting leads, a configuration that remains less explored in the nonlinear regime. We focus on the T-shaped geometry because in this configuration QD$_2$ induces the Fano-Andreev effect, giving rise to Fano-type interference or hybridized structure on QD$_1$ spectrum~\cite{Baranski2011,Calle2013,Calle2017,Baranski2020,Gorski2021,Gonzalez2024,Gonzalez2026}. The interplay between the Fano-Andreev effect and the internal screening potential, which shifts the lead-coupled central dot level under nonequilibrium conditions, plays a crucial role in nonlinear transport. We first analyze the limiting cases of purely voltage bias and purely thermal driving, and then address the weakly nonlinear regime under simultaneous voltage and thermal biases. Our analysis is based on the Keldysh nonequilibrium Green’s function formalism, and the charge redistribution and screening effects due to Coulomb interaction on the center dot is incorporated through characteristic internal potentials $U(V,\theta)$ within a gauge-invariant nonlinear transport framework~\cite{Christen1996,Wang2001,Pilgram2002,Hwang2015,Meair2013,Sanchez2013}. The rest of the paper is organized as follows. In Sec.~\ref{sec:2}, we introduce the model Hamiltonian and outline the theoretical framework. Section~\ref{sec:3} presents the numerical results for the nonlinear transport properties and discusses their physical implications. Finally, Sec.~\ref{sec:4} summarizes the main results and concludes the paper.
\section{Model Hamiltonian and theoretical description}
\label{sec:2}
\paragraph*{\textbf{Bogoliubov-transformed Hamiltonian and Green's function:}}  The T-shaped double quantum dot coupled to normal and superconducting leads is described by an extended Anderson Hamiltonian, with the superconducting lead represented by the Bogoliubov-transformed BCS mean-field Hamiltonian~\cite{Verma2022}. The total Hamiltonian is given by,
\begin{equation} \label{eq:1}
\begin{aligned}
H &=  \sum_{k,\sigma}\epsilon_{k,N}c^\dagger_{k\sigma,N}c_{k\sigma,N}
+\sum_{k,\sigma}E_{k}\gamma^\dagger_{k\sigma}\gamma_{k\sigma}
+\sum_{j=1,2}\sum_{\sigma} (\epsilon_{j}-e\,U_j)\,n_{j\sigma}+\sum_{\sigma}t\,(d^\dagger_{1\sigma}d_{2\sigma}+\text{H.c.})\\[4pt]
&+\sum_{k\sigma}(\mathcal{V}_{k,N} d^\dagger_{1\sigma}c_{k\sigma,N}+\text{H.c.})
+\sum_{k\sigma}(\mathcal{V}_{k,S}u^\ast_k d^\dagger_{1\sigma}\gamma_{k\sigma}+\text{H.c.})+\sum_{k}\Big[\mathcal{V}^{\ast}_{k,S}v_k(d^\dagger_{1\uparrow}\gamma^\dagger_{-k\downarrow}-d^\dagger_{1\downarrow}\gamma^\dagger_{k\uparrow})+\text{H.c.}\Big].
\end{aligned}
\end{equation}
The first term describes the normal metallic reservoir (N), where $c_{k\sigma,N}$ ($c^\dagger_{k\sigma,N}$) annihilates (creates) an electron with spin $\sigma$ and momentum $\vec{k}$, and $\epsilon_{k,N}$ denotes the corresponding single-particle energy measured relative to the chemical potential. The second term describes the superconducting reservoir (S), where $\gamma_{k\sigma}$ ($\gamma^\dagger_{k\sigma}$) denotes the annihilation (creation) operator for Bogoliubov quasiparticles with spin $\sigma$, wave vector $\vec{k}$, and energy $E_k = \sqrt{\epsilon_{k,S}^2 + |\Delta|^2}$, where $\epsilon_{k,S}$ is the single-particle energy in the superconducting lead measured relative to the chemical potential, and $\Delta$ is the superconducting energy gap.\\
The third term describes the two quantum dots with energy levels $\epsilon_j$ ($j=1,2$), where $d_{j\sigma}$ ($d^\dagger_{j\sigma}$) annihilates (creates) an electron on dot $j$, and $n_{j\sigma}=d^\dagger_{j\sigma}d_{j\sigma}$ is the corresponding occupation number operator. Electron-electron interaction effects are incorporated through an internal electrostatic potential $U_1=U(V,\theta)$ acting on the lead-coupled central dot (QD$_1$) within the mean-field approximation. This potential accounts for the screening induced by nonequilibrium charge redistribution on the lead-coupled dot, providing an effective description of the Coulomb interaction for metallic-dots, where screening is efficient~\cite{Hwang2015}. Consequently, the energy level of QD$_1$ is renormalized as $\tilde{\epsilon}_1=\epsilon_1-e\,U(V,\theta)$. The side-coupled dot (QD$_2$) level $\epsilon_2$ remains unchanged ($U_2 = 0$), as it is not directly coupled to the leads.\\
The fourth term accounts for interdot tunneling with amplitude $t$, which couples QD$_1$ and QD$_2$ and gives rise to quantum interference of the Fano type. The remaining terms describe the tunneling between QD$_1$ and leads, with $\mathcal{V}_{k,\alpha}$ denoting the tunneling amplitude between the QD$_1$ and the $\alpha$-lead ($\alpha \in \mathrm{N}, \mathrm{S}$). The Bogoliubov coefficients $u_k$ and $v_k$ are given by
\begin{equation} \label{eq:2}
|u_k|^2=\frac{1}{2}\left(1+\frac{\epsilon_{k,S}}{\sqrt{\epsilon_{k,S}^2+|\Delta|^2}}\right), 
\qquad 
|v_k|^2=\frac{1}{2}\left(1-\frac{\epsilon_{k,S}}{\sqrt{\epsilon_{k,S}^2+|\Delta|^2}}\right).
\end{equation}
Within the wide flat-band limit, assuming momentum-independent tunneling amplitudes $\mathcal{V}_{\alpha}$. The tunnel coupling (tunneling rates) between the QD$_1$ and lead $\alpha$ is $\Gamma_{\alpha} = 2\pi |\mathcal{V}_{\alpha}|^2 \rho_{\alpha}$,
where $\rho_{\alpha}$ is the (normal-state) density of states of lead $\alpha$, assumed to be constant near the Fermi level. Since QD$_2$ is not directly coupled to the leads, it does not acquire any direct tunneling-induced broadening.\\
Next, the retarded Green’s function of the system $\mathbf{G}^r(\omega)$ is obtained using equation of motion method and Dyson equation, and its explicit $4\times4$ Nambu matrix structure-incorporating superconducting pairing, interdot tunneling and interaction-induced level renormalization is given as
\begin{equation} \label{eq:3}
\mathbf{G}^r(\omega)
=
\begin{pmatrix}
\omega-\tilde{\epsilon}_1 + \dfrac{i}{2}\Gamma_N + \dfrac{i}{2}\Gamma_S \beta(\omega) & t & -\dfrac{i}{2}\Gamma_S\,\dfrac{\Delta}{|\omega|}\,\beta(\omega) & 0 \\[10pt]

t & \omega-\epsilon_2 & 0 & 0 \\[10pt]

-\dfrac{i}{2}\Gamma_S\,\dfrac{\Delta}{|\omega|}\,\beta(\omega) & 0 &
\omega+\tilde{\epsilon}_1 + \dfrac{i}{2}\Gamma_N + \dfrac{i}{2}\Gamma_S \beta(\omega) & -t \\[10pt]

0 & 0 & -t & \omega+\epsilon_2
\end{pmatrix}^{-1},
\end{equation}
where $\omega \rightarrow \omega + i0^+$. The renormalized energy level of QD$_1$ is given by $\tilde{\epsilon}_1=\epsilon_1-e\,U(V,\theta)$, where $U(V,\theta)$ is the bias- and temperature-induced internal potential. Here, $\Gamma_N$ and $\Gamma_S$ represent the tunneling coupling strengths between QD$_1$ and the normal and superconducting leads, respectively. The function $\beta(\omega)$ is defined as
\begin{equation} \label{eq:4}
\beta(\omega) =
\frac{\omega}{i\sqrt{\Delta^2 - \omega^2}}\,\Theta(\Delta - |\omega|)
+ \frac{|\omega|}{\sqrt{\omega^2 - \Delta^2}}\,\Theta(|\omega| - \Delta),
\end{equation}
with $\Theta(x)$ being the Heaviside step function. The explicit analytical expressions for the individual Green's function components associated with QD$_1$, are given in Appendix~\ref{App:A}.\\

\paragraph*{\textbf{Nonlinear transport theory:}}
Let the system is under the influence of finite voltage $V=(\mu_N-\mu_S)/e$ (say $\mu_N=eV$ and $\mu_S=0$) and temperature gradient $T_N-T_S=\theta$ (say $T_N=T+\theta$ and $T_S=T$). The net charge current is calculated from the time evolution of the electron number operator of the normal reservoir~\cite{Meir1992,Cuevas1996,Sun1999,Hwang2016b}. Using the Heisenberg equation of motion, the charge current is expressed as
\begin{equation} \label{eq:5}
I_C(V,\theta) = -\frac{ie}{\hbar} \langle [H, \sum_{k\sigma} c^{\dagger}_{k\sigma,N} c_{k\sigma,N}] \rangle.
\end{equation}
Within the nonequilibrium Keldysh Green's function formalism, the charge current $I_C(V,\theta)$ flowing across the system (under simultaneous voltage and thermal biases) from the normal to the superconducting reservoir can be written as
\begin{equation}\label{eq:6}
I_C(V,\theta) = I_A(V,\theta)+I_{QP}(V,\theta);
\end{equation}
with
\begin{equation} \label{eq:7}
I_A(V,\theta) = \frac{2e}{h} \int{d\omega\;T_A(\omega)\;\left[f_N(\omega-eV, T+\theta)-f_N(\omega+eV, T+\theta)\right]}
\end{equation}
\begin{equation} \label{eq:8}
I_{QP}(V,\theta) = \cfrac{2e}{h} \int{T_{QP}(\omega)\;d\omega\;\left[f_N(\omega-eV, T+\theta)-f_S(\omega, T)\right]} 
\end{equation}
where $I_A(V,\theta)$ and $I_{QP}(V,\theta)$ denote the Andreev and quasiparticle contributions to the charge currents, respectively. The Fermi-Dirac distribution function of the leads is given by $f_{\alpha\in N,S}(\omega\mp\mu_{\alpha}, T_{\alpha})=\left[{exp((\omega\mp\mu_{\alpha})/k_BT_{\alpha})+1}\right]^{-1}$, where $T_\alpha$ and $\mu_\alpha$ are the temperature and chemical potential of reservoir $\alpha$, respectively. The chemical potentials $\pm \mu_\alpha$ are measured with respect to the Fermi level, which is taken as $E_F = 0$.\\
The derivation of the Andreev contribution from the charge-current formula within the nonequilibrium Green's-function formalism is
standard and has been presented in detail in
Refs.~\cite{Cuevas1996,Sun1999}. For the present geometry, in which the normal reservoir is coupled only to QD$_1$, the transmission coefficients for Andreev reflection and quasiparticle transport are
given by~\cite{Baranski2011,Calle2017,Hwang2015,Wang2001}
\begin{equation} \label{eq:9}
T_A(\omega)=\Gamma_N^2 \left|G_{13}^r(\omega)\right|^2,
\end{equation}
\begin{equation} \label{eq:10}
T_{QP}(\omega) =
\dfrac{\Gamma_N \Gamma_S |\omega|}{\sqrt{\omega^2-\Delta^2}}\,
\theta(|\omega|-\Delta)
\left[
|G_{11}^r(\omega)|^2
+ |G_{13}^r(\omega)|^2
- \dfrac{2\Delta}{|\omega|}
\mathrm{Re}\left\{G_{11}^r(\omega)\, G_{13}^a(\omega)\right\}
\right].
\end{equation}
Here, $G_{11}^r(\omega)$ and $G_{13}^r(\omega)$ denote the normal (electron-electron) and anomalous (electron-hole) components of the retarded Green's function of QD$_1$, respectively, while $G_{13}^a(\omega)$ denotes the corresponding anomalous (electron-hole) component of the advanced Green's function, as defined in Appendix~\ref{App:A}.\\
Within the Keldysh formalism, the lesser Green's function is used to evaluate the nonequilibrium charge on the lead-coupled central quantum dot, QD$_1$. It is obtained from the following Keldysh equation~\cite{Cuevas1996,Wang2001,Sun1999}
\begin{equation} \label{eq:11}
\mathbf{G}^{<}(\omega)=\mathbf{G}^{r}(\omega)\,\mathbf{\Sigma}^{<}(\omega)\,\mathbf{G}^{a}(\omega).
\end{equation}
The advanced Green's function matrix is $\mathbf{G}^{a}(\omega) = \left[\mathbf{G}^{r}(\omega)\right]^{\dagger}$ and the lesser self energy matrix is obtained by
\begin{equation} \label{eq:12}
{\bf{\Sigma}}^{<}(\omega)= -\sum_{\alpha\in N,S}\left[{\bf{\Sigma}}^{r}_{\alpha}-{\bf{\Sigma}}^{a}_{\alpha}\right]f_{\alpha}(\omega-\mu_{\alpha})
\end{equation}
Now, using the expressions for the retarded and advanced self-energies from Eq.~\ref{eq:A7}, we obtain
\begin{equation} \label{eq:13}
\mathbf{\Sigma}^{<}(\omega)=
\begin{pmatrix}
\Sigma^{<}_{11}(\omega) & 0 & \Sigma^{<}_{13}(\omega) & 0 \\[8pt]

0 & 0 & 0 & 0 \\[8pt]

\Sigma^{<}_{31}(\omega) & 0 & \Sigma^{<}_{33}(\omega) & 0 \\[8pt]

0 & 0 & 0 & 0
\end{pmatrix}
\end{equation}
with
\begin{align}
\Sigma^{<}_{11}(\omega) &= i\Gamma_{N}f_N(\omega-eV)+\cfrac{i\Gamma_S|\omega|}{\sqrt{\omega^2-\Delta^2}} \theta(|\omega|-\Delta)f_S(\omega) \label{eq:14}\\[8pt]
\Sigma^{<}_{13}(\omega) &= \Sigma^{<}_{31}(\omega)=\cfrac{-i\Gamma_{S}\Delta}{\sqrt{\omega^2-\Delta^2}}\theta(|\omega|-\Delta)f_S(\omega) \label{eq:15} \\[8pt]
\Sigma^{<}_{33}(\omega) &= i\Gamma_{N}f_N(\omega+eV)+\cfrac{i\Gamma_S|\omega|}{\sqrt{\omega^2-\Delta^2}} \theta(|\omega|-\Delta)f_S(\omega) \label{eq:16}
\end{align}
From Eq.~\ref{eq:11}, we get the lesser Green's function for electrons on the QD$_1$ as
\begin{equation} \label{eq:17}
\begin{aligned}
G^{<}_{11}(\omega) &=   i\Gamma_{N}f_N(\omega-eV)|G_{11}^{r}(\omega)|^2+
i\Gamma_{N}f_N(\omega+eV)|G_{13}^{r}(\omega)|^2\\
&+\frac{i\Gamma_S|\omega|}{\sqrt{\omega^2-\Delta_2}}\;\theta(|\omega|-\Delta)f_S(\omega)\,\left[|G_{11}^{r}(\omega)|^2+|G_{13}^{r}(\omega)|^2
-\frac{2\Delta}{|\omega|} Re\left\{G_{11}^{r}(\omega)\times G_{13}^{a}(\omega)\right\}\right]
\end{aligned}
\end{equation}
In hybrid normal-superconducting systems, $G_{11}^{<}(\omega)$ contains contributions from both the normal electronic and anomalous particle-hole sectors. Within the superconducting gap, $|\omega|<\Delta$, transport is governed by Andreev bound states, and the anomalous component $|G_{13}^{r}(\omega)|^{2}$ describes the particle-hole conversion responsible for the subgap Andreev current. However, charge occupation and screening responses generally depend on both the normal electronic component, $G_{11}^{r}(\omega)$, and the anomalous component, $G_{13}^{r}(\omega)$. For $|\omega|>\Delta$, the superconducting quasiparticle continuum becomes accessible, and $G_{11}^{<}(\omega)$ is dominated by the normal electronic contribution to the dot occupation, represented by $|G_{11}^{r}(\omega)|^{2}$ weighted by the nonequilibrium lead distribution functions, together with the quasiparticle contribution arising from the superconducting continuum.\\
Eq.~\ref{eq:17} allows us to determine the charge and entropic injectivities, as well as the Lindhard screening function, which together govern the bias-induced variation of the internal potential. These quantities are evaluated at equilibrium using the retarded Green’s functions of the system, as detailed in Appendix~\ref{App:B}. To leading order in the applied voltage and thermal biases, the resulting shift in the internal electrostatic potential is expressed in terms of the characteristic potentials as $\delta U(V,\theta)=u_N V+z_N\theta$. The corresponding Green’s functions are subsequently used to compute the current according to Eqs.~\ref{eq:6}, \ref{eq:7} and \ref{eq:8}.\\

\paragraph*{\textbf{Pure voltage bias (isothermal case):}}  
For purely electrical driving ($V \neq 0$ and $\theta = 0$), transport is characterized by the differential (nonlinear) electrical conductance, defined as
\begin{equation} \label{eq:18}
G(V)
=
\left.\frac{\partial I_C(V,0)}{\partial V}\right|_{\theta=0}.
\end{equation}
The total conductance naturally separates into Andreev and quasiparticle contributions,
\begin{equation} \label{eq:19}
G(V) = G_A(V) + G_{QP}(V),
\end{equation}
with
\begin{equation} \label{eq:20}
G_A(V) = \left.\frac{\partial I_A(V)}{\partial V}\right|_{\theta=0}, 
\qquad
G_{QP}(V) = \left.\frac{\partial I_{QP}(V)}{\partial V}\right|_{\theta=0}.
\end{equation}
In this regime, both quasiparticle and Andreev processes contribute to the charge current. In particular, subgap transport ($|eV| < \Delta$) is dominated by Andreev reflection, which enables finite current via electron-hole conversion, while quasiparticle contributions become significant only for $|eV| > \Delta$.\\

\paragraph*{\textbf{Pure thermal driving (isoelectric case):}}
At zero voltage bias and finite thermal bias ($V = 0$ and $\theta \neq 0$), the Andreev current vanishes identically for any thermal bias, $I_A(0,\theta)=0$, reflecting particle-hole symmetry of subgap Andreev reflection. Therefore, under pure thermal driving, the thermoelectric response is entirely determined by the quasiparticle contribution. The nonlinear Seebeck coefficient is defined under open-circuit conditions, where
$V_{\mathrm{th}}$ denotes the thermovoltage generated to maintain vanishing total charge current,
\begin{equation} \label{eq:21}
S_{\mathrm{NL}}(\theta)
=
-
\left.
\frac{V_{\mathrm{th}}}{\theta}
\right|_{I_C(V_{\mathrm{th}},\theta)=0}.
\end{equation}

\paragraph*{\textbf{Simultaneous voltage and thermal bias (cross thermoelectric coupling):}}
When a finite voltage and thermal bias are applied simultaneously ($V\neq0$ and $\theta\neq0$), the electrical and thermal drivings become coupled, giving rise to a nonlinear cross thermoelectric effect. In the weakly nonlinear regime, the leading-order mixed voltage-temperature contribution to the Andreev current in Eq.~\ref{eq:7} is approximately given by~\cite{Hwang2015}
\begin{equation}  \label{eq:22}
\delta I_A(V,\theta) =  I_A(V,\theta) -  I_A(V,0)
\simeq
M_{1}V\theta,
\end{equation}
where $M_{1}$ is Andreev contribution to the cross thermoelectric coupling coefficient. It is defined as
\begin{equation} \label{eq:23}
M_1
=
\left.
\frac{\partial^2 I_A}{\partial V\,\partial\theta}
\right|_{\mathrm{eq}},
\end{equation}
which quantifies the leading-order nonlinear coupling between the electrical and thermal biases in the vicinity of equilibrium and is given by
\begin{equation} \label{eq:24}
M_1
=
\frac{4e^2}{h}
\int d\omega
\left(
-\frac{\partial f}{\partial\omega}
\right)
\left[
z_N\frac{\partial T_A}{\partial U}
+
\frac{\omega}{T}
\frac{\partial T_A}{\partial \omega}
\right]_{\mathrm{eq}}.
\end{equation}
The first term represents the interaction-induced screening response through the characteristic thermal potential $z_N$, whereas the second term is the direct noninteracting contribution arising from the energy dependence of the Andreev transmission. Consequently, the sign of $M_1$ determines whether the magnitude of the Andreev current is enhanced or suppressed by a small thermal bias at finite voltage.

\section{Result and discussion}\label{sec:3}
In this section, we present numerical results for the weakly nonlinear electric and thermoelectric response. We analyze three regimes discussed in the previous section: (i) under a purely voltage bias, subgap transport is governed by Andreev processes; (ii) under a purely thermal bias, the response arises solely from above-gap quasiparticle transport, generating thermovoltage and thermopower, while subgap Andreev processes do not contribute due to intrinsic particle-hole symmetry; and (iii) simultaneous voltage and thermal biases leads to a finite nonlinear thermoelectric response driven by Andreev processes via cross thermoelectric coupling. Throughout this work, the superconducting gap $\Delta$ is taken as the unit of energy.
\begin{figure*}[!htb]
\centering
\includegraphics
[width=0.45\hsize]
{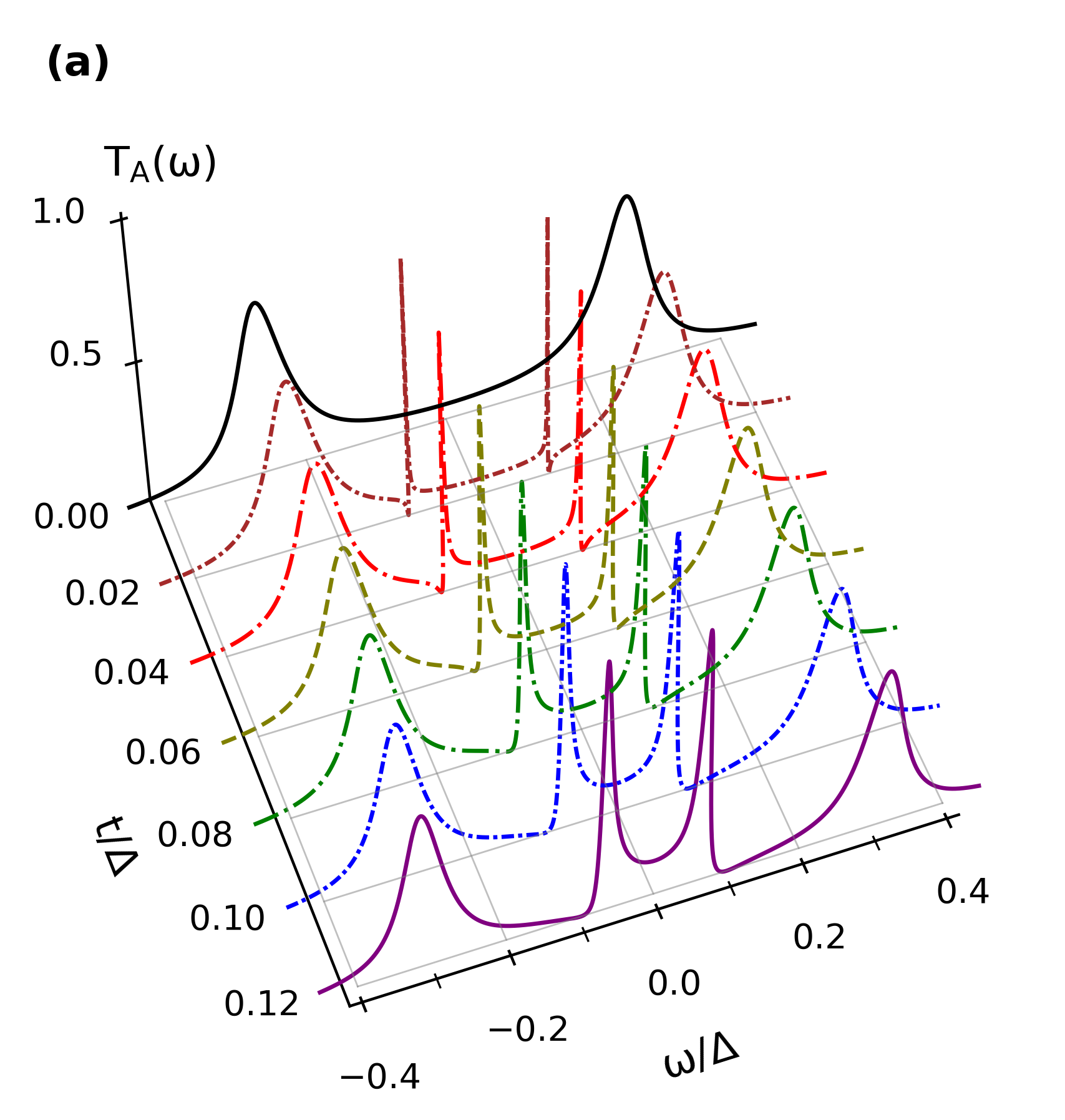}
\includegraphics
[width=0.45\hsize]
{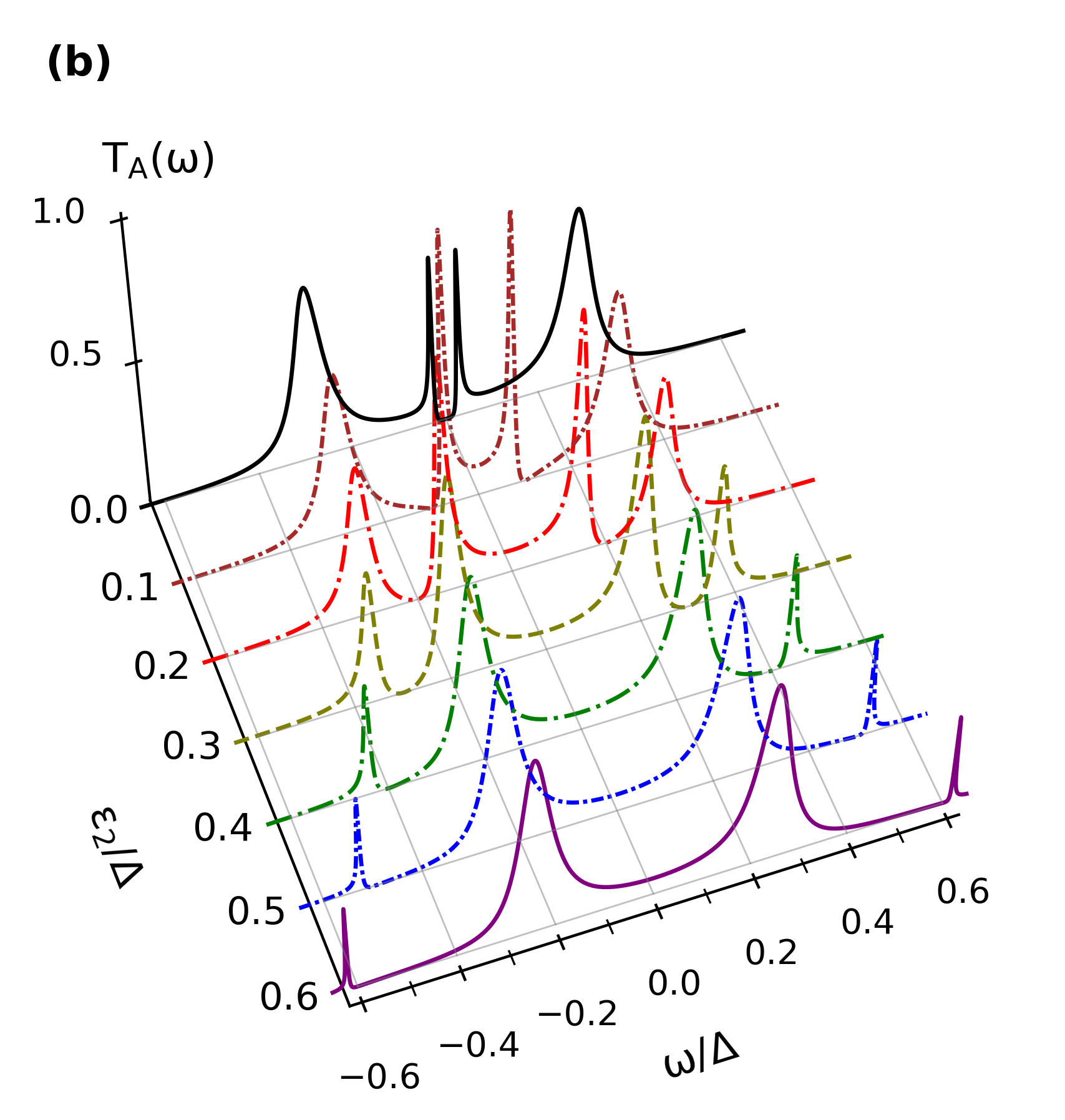}
\caption{
Andreev transmission coefficient $T_A(\omega)$ at equilibrium as a function of energy $\omega$, showing the evolution of the spectral signatures of the Fano-Andreev effect. (a) Evolution with interdot tunneling amplitude $t$ for fixed QD$_2$ energy $\epsilon_2 = 0.1\Delta$. (b) Evolution with QD$_2$ energy level $\epsilon_2$ for fixed interdot coupling $t = 0.1\Delta$. The remaining parameters are $\epsilon_1 = 0.2\Delta$, $\Gamma_N = 0.1\Delta$, and $\Gamma_S = 0.5\Delta$.}
\label{fig:2}
\end{figure*}\\
Before analyzing the nonlinear transport characteristics, we first discuss the equilibrium spectral properties of the system. Fig.~\ref{fig:2}(a) shows the Andreev transmission coefficient $T_A(\omega)$ as a function of energy $\omega$ for different values of the interdot coupling $t$ ranging from the interferometric to molecular regime. In equilibrium and for $t=0$ (i,e., when QD$_2$ is effectively decoupled), the subgap response is governed by the proximity-induced pairing on QD$_1$, which gives rise to coherent Andreev reflection. Consequently, $T_A(\omega)$ exhibits pronounced peaks at energies corresponding to the Andreev bound states. For finite interdot coupling, the interplay between Fano interference and Andreev reflection gives rise to the Fano-Andreev effect. In the weak-coupling regime ($t \ll \Gamma_N$), QD$_2$ introduces an energy-dependent self-energy, $\Sigma_{\mathrm{QD}_2}(\omega)\sim t^2/(\omega-\epsilon_2)$, which produces characteristic Fano-type lineshapes (dip-peak structures) around $\omega \approx \pm \epsilon_2$. These features originate from quantum interference between the direct transmission path through QD$_1$ and the indirect path via QD$_2$. Notably, the resonances at $\omega \approx \pm \epsilon_2$ have distinct microscopic origins: the feature near $\omega \approx \epsilon_2$ arises from conventional Fano interference due to direct scattering through QD$_2$, whereas the structure near $\omega \approx -\epsilon_2$ originates from anomalous Fano interference induced by particle-hole mixing due to superconducting pairing~\cite{Baranski2011,Baranski2020,Gorski2021}. While well-defined Fano-type lineshapes appear for weak interdot coupling $t$ in the interferometric regime, they gradually evolve into more symmetric hybridized quasiparticle peaks for larger interdot coupling ($t \approx \Gamma_N$), corresponding to the molecular regime. Physically, this reflects the leakage of superconducting correlations into QD$_2$, which feed back onto QD$_1$ and lead to the formation of hybridized Andreev-Fano states.\\ 
Fig.~\ref{fig:2}(b) shows $T_A(\omega)$ as a function of $\omega$ for different values of QD$_2$ energy level $\epsilon_2$. As discussed above, characteristic Fano-type lineshapes appear at $\omega \approx \pm \epsilon_2$. For $\epsilon_2 = 0$, a pronounced antiresonance occurs at $\omega = 0$, indicating destructive interference at the Fermi level. In this case, the zero-energy spectral weight is suppressed and redistributed to finite energies. When $\epsilon_2$ lies close to the Fermi level, the low energy Andreev transmission is strongly influenced.  For relatively larger $|\epsilon_2|$, the signatures of the Fano-Andreev effect shifts to finite energies, leading to a recovery of low energy Andreev transmission, while Fano resonances persist near $\omega \approx \pm \epsilon_2$.\\

\paragraph*{\textbf{$\mathrm{(i)}$ Pure voltage bias (isothermal case):}}
We now consider weakly nonlinear transport under a pure voltage bias, where a finite voltage $V$ is applied across the junction while both leads are maintained at the same temperature. The applied voltage modifies the occupation of QD$_1$, leading to nonequilibrium charge accumulation. The resulting electrostatic screening generates an internal potential shift, $\delta U(V)$, which renormalizes the QD$_1$ energy level. The response is approximately antisymmetric and linear in $V$, reflecting the leading-order behaviour $\delta U \simeq u_N V$ in the weakly nonlinear regime. In the limit of weak capacitive coupling ($C \to 0$), the characteristic potential reduces to $u_N \approx D_N/\Pi$, where $D_N$ is the charge injectivity describing the voltage-induced change in the charge accumulated on QD$_1$, and $\Pi$ is the Lindhard function characterizing the electronic screening response to the resulting electrostatic perturbation. With increasing interdot coupling $t$, the system enters the molecular regime, leading to the formation of hybridized Andreev-Fano states on QD$_1$. These states redistribute the relative spectral weights of the normal electron-like and anomalous particle-hole contributions that determine the charge injectivity $D_N$ in Eq.~\ref{eq:B10}.
\begin{figure*}[!htb]
\centering
\includegraphics
[width=0.39\hsize]
{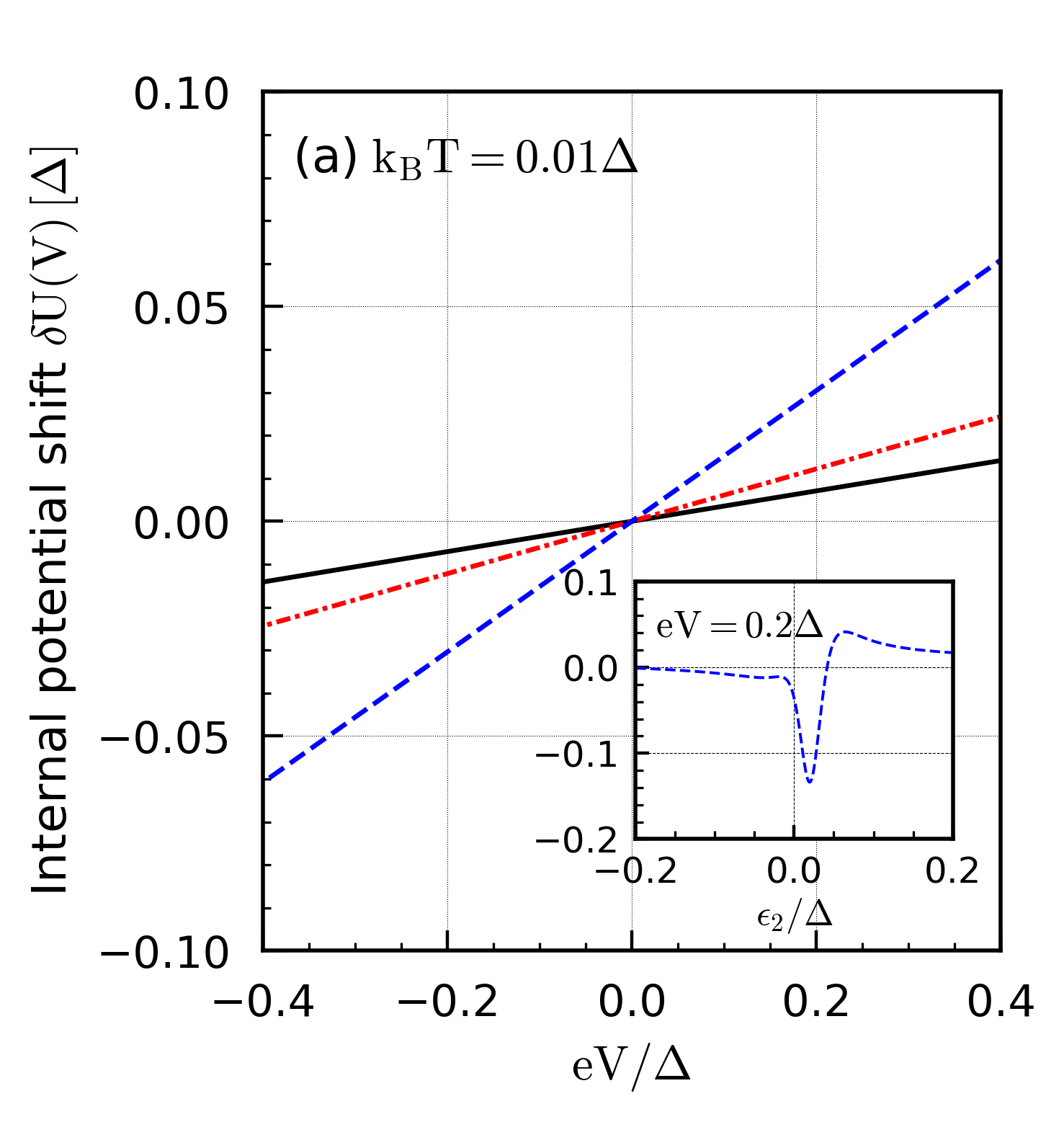}
\includegraphics
[width=0.39\hsize]
{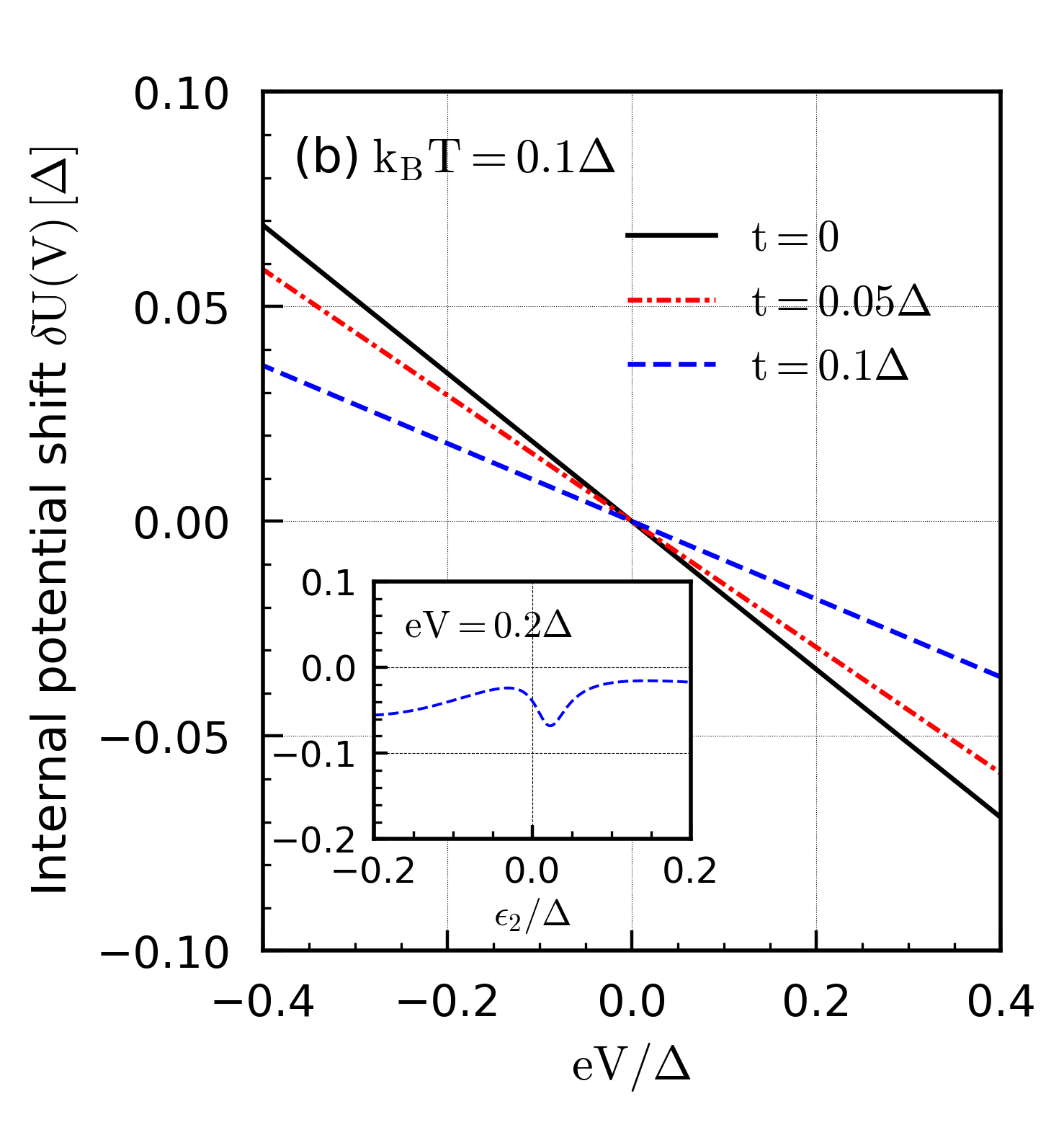}
\caption{
Internal-potential shift, $\delta U(V)$, obtained within the weakly nonlinear characteristic-potential approach, as a function of the applied bias voltage $V$ for different interdot coupling strengths $t$: (a) $k_BT=0.01\Delta$ and (b) $k_BT=0.1\Delta$, with $\epsilon_2=0.1\Delta$. The insets show $\delta U$ as a function of $\epsilon_2$ for $t=0.1\Delta$ at $eV=0.2\Delta$. Other parameters are $\epsilon_1=0.2\Delta$, $\Gamma_N=0.1\Delta$, and $\Gamma_S=0.5\Delta$.
}
\label{fig:3}
\end{figure*}
\begin{figure*}[!htb]
\centering
\includegraphics
[width=0.4\hsize]
{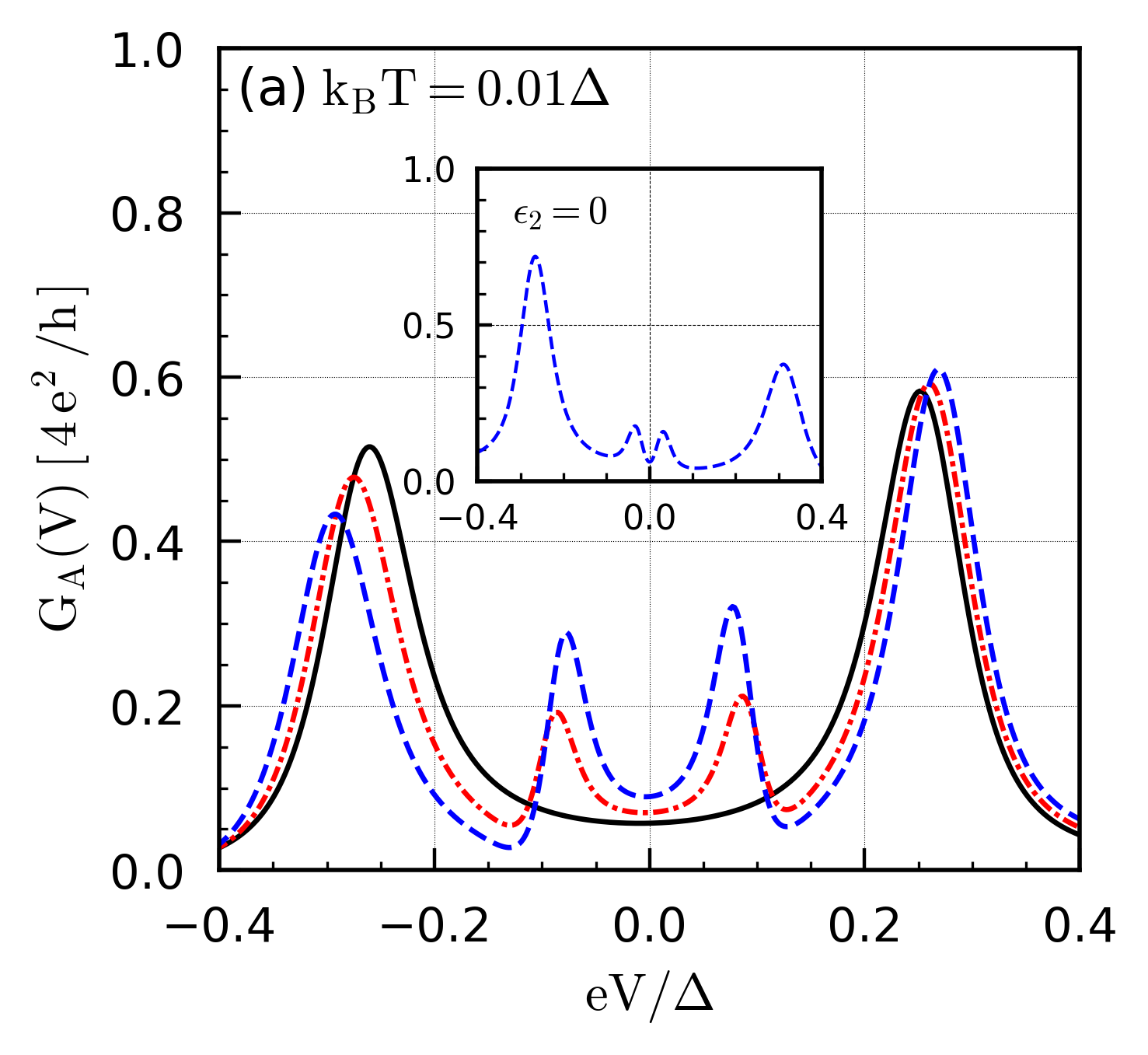}
\includegraphics
[width=0.4\hsize]
{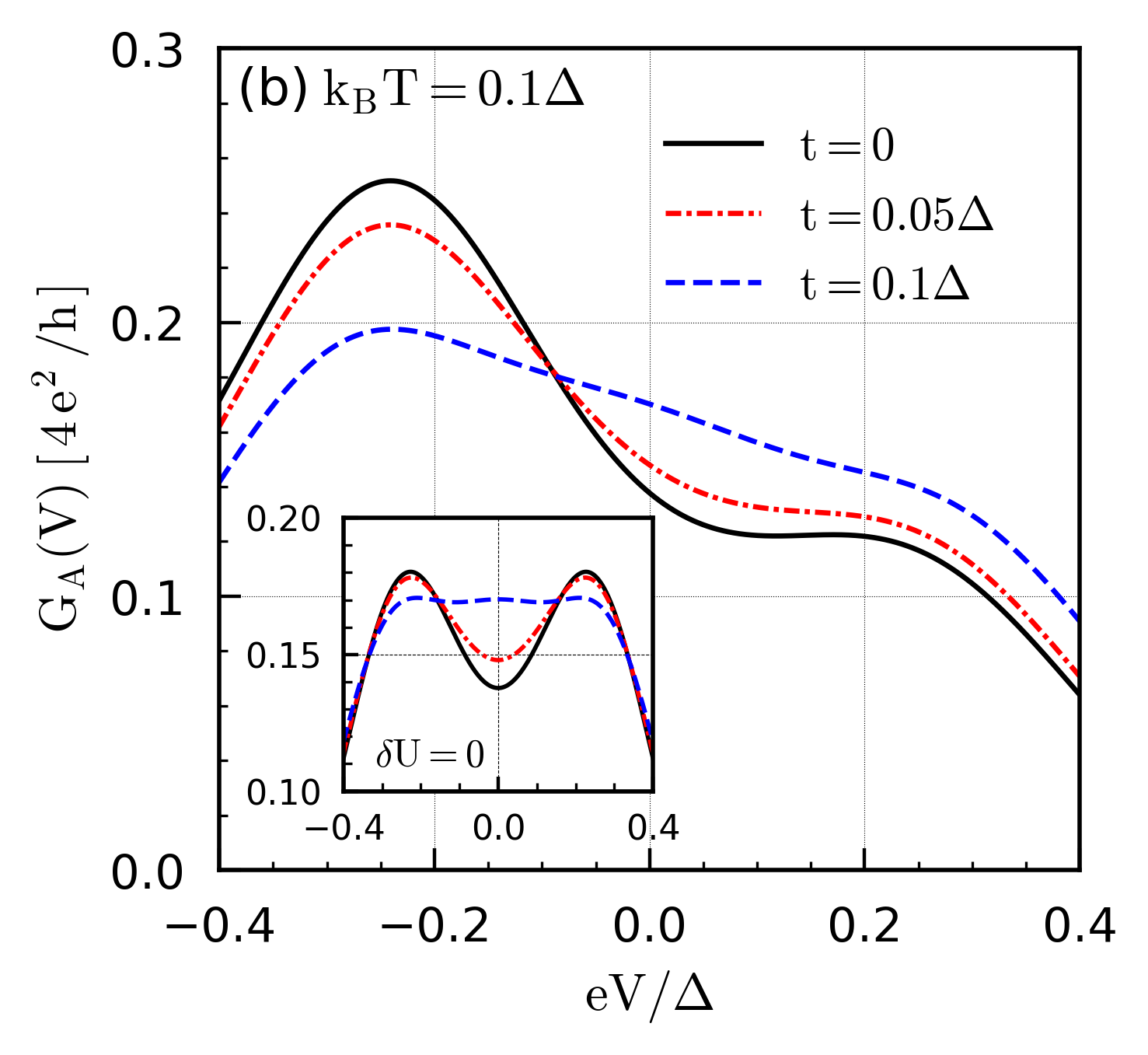}
\caption{Andreev differential conductance, $G_A(V)=dI_A/dV$, as a function of the applied bias voltage $V$ for different interdot couplings $t$ with $\epsilon_2=0.1\Delta$: (a) $k_BT=0.01\Delta$ and (b) $k_BT=0.1\Delta$. The remaining parameters are $\epsilon_1=0.2\Delta$, $\Gamma_N=0.1\Delta$, and $\Gamma_S=0.5\Delta$. The inset in panel (a) shows the corresponding conductance for $\epsilon_2=0$ at $t=0.1\Delta$, illustrating the conductance when the QD$_2$ level is aligned with the Fermi level. The inset in panel (b) shows the corresponding conductance in the absence of the internal potential shift, i.e., $\delta U=0$.}
\label{fig:4}
\end{figure*}\\
Figure~\ref{fig:3}(a) shows the leading-order nonequilibrium internal potential shift as a function of the applied voltage for different interdot couplings $t$ at the low background temperature $k_BT=0.01\Delta$, where the charge injectivity is dominated by the low-energy spectral weight near the Fermi level. For $\epsilon_1$ close to the Fermi level and a small positive QD$_2$ level ($\epsilon_2\simeq0.1\Delta$), increasing the interdot coupling strengthens the mixing between the proximity-induced Andreev states on QD$_1$ and QD$_2$ level. The resulting Andreev-Fano hybridization redistributes the normal electron-like and anomalous particle-hole spectral weights within the low-energy window relevant to the charge injectivity. In this regime, the relative anomalous contribution entering Eq.~\ref{eq:B10} increases, resulting in a larger $D_N$. The corresponding enhancement of the ratio $u_N=D_N/\Pi$ produces the steeper slope of $\delta U(V)$ with increasing $t$. The inset shows $\delta U$ as a function of the QD$_2$ energy level $\epsilon_2$ for $t=0.1\Delta$ at fixed positive voltage. A pronounced asymmetric dip appears near $\epsilon_2 \rightarrow 0^+$, which is a hallmark of destructive Fano interference. At this condition, the interference suppresses the low-energy spectral weight on QD$_1$, leading to a strong reduction of the anomalous contribution, while the normal contribution remains finite. Likewise, tuning $\epsilon_1$ away from the Fermi level decreases the slope and eventually changes its sign (not shown), reflecting the dominance of the normal contribution to the injectivity.\\
Figure~\ref{fig:3}(b) shows the internal potential shift $\delta U(V)$ at the higher background temperature $k_BT=0.1\Delta$. In contrast to the low-temperature case, thermal broadening extends the energy window contributing to the charge injectivity, allowing higher-energy electron-like states to contribute significantly. As a result, the normal electron-like contribution increases more rapidly than the anomalous particle-hole contribution, thereby reducing the relative influence of the latter on the total injectivity $D_N$. Although increasing the interdot coupling $t$ continues to enhance the hybridization of the Andreev-Fano states, the thermally enhanced normal contribution dominates the screening response, leading to a reduction in the characteristic potential $u_N$. Consequently, the magnitude of the slope of $\delta U(V)$ decreases progressively with increasing $t$. The inset shows $\delta U$ as a function of QD$_2$ energy level $\epsilon_2$ for $t=0.1\Delta$ at a fixed positive bias voltage. The asymmetric Fano feature near $\epsilon_2\rightarrow0^+$ remains visible but is broadened by thermal effects, reflecting the reduced influence of Fano-Andreev interference on the overall screening response.\\ 
Figure~\ref{fig:4}(a) shows the Andreev differential conductance, $G_A(V)$, calculated within the first-order characteristic-potential approximation as a function of the applied bias voltage $V$ at the low background temperature $k_BT=0.01\Delta$. For $t=0$, two broad conductance peaks associated with the conventional Andreev bound states are observed. Finite interdot coupling modifies these states through hybridization with QD$_2$ level, resulting in additional low-bias substructures. This low-bias conductance increases with interdot coupling $t$ due to the enhanced proximity-induced pairing associated with the Fano-Andreev effect. The internal potential shift $\delta U(V)$ introduces a bias asymmetry in the Andreev differential conductance through the bias-dependent renormalization of the QD$_1$ energy level. For $\epsilon_2=0.1\Delta$ (main panel), the sign of $\delta U(V)$ under bias reversal shifts the effective level $\tilde{\epsilon}_1$ alternately toward and away from the Fermi level. Consequently, the positive-bias Andreev-bound-state peak is slightly enhanced, whereas the corresponding negative-bias peak is suppressed. In contrast, the inset shows the conductance for $\epsilon_2=0$ at $t=0.1\Delta$, where destructive Fano interference reverses the sign of the internal potential relative to the main panel. Consequently, the effective QD$_1$ level is shifted in the opposite direction, suppressing the positive-bias Andreev-bound-state peak while enhancing the corresponding negative-bias peak. Thus, nonlinear screening modifies only the bias asymmetry, whereas the low-bias conductance is governed primarily by the underlying Fano-Andreev resonance structure.\\
Figure~\ref{fig:4}(b) shows the corresponding conductance at the higher background temperature $k_BT=0.1\Delta$. Thermal broadening suppresses the sharp Fano-Andreev resonances, resulting in a much smoother conductance spectrum. In the absence of the internal potential shift, shown in the inset, the conductance remains nearly symmetric with respect to bias reversal, and its evolution with increasing interdot coupling is governed primarily by the equilibrium Fano-Andreev interference. In contrast, when the internal potential shift is included, increasing $t$ progressively suppresses the negative-bias conductance while enhancing the conductance over a broad positive-bias region, giving rise to a pronounced bias asymmetry. As discussed in Fig.~\ref{fig:3}(b), thermal broadening alters the relative balance between the normal and anomalous contributions to the injectivity, with the normal contribution becoming comparable to and eventually exceeding the anomalous contribution. Consequently, the bias-dependent renormalization of the QD$_1$ energy level has a more pronounced effect on the thermally broadened conductance than at low temperature. Thus, although thermal broadening weakens the intrinsic Fano-Andreev interference, it makes the influence of nonlinear screening on the conductance much more clearly distinguishable through the pronounced bias asymmetry.\\
\paragraph*{\textbf{$\mathrm{(ii)}$ Pure thermal driving (isoelectric case):}} Next, we consider the isoelectric case, $V=0$, where the system is driven solely by a thermal bias $\theta$. Since the thermoelectric response under pure thermal driving is governed by quasiparticle transport above the superconducting gap, we assume symmetric couplings to the normal and superconducting leads, $\Gamma_N=\Gamma_S$. In the limit $C\to 0$, the characteristic thermal potential reduces to $z_N=\widetilde{D}_N/\Pi$, where $\widetilde{D}_N$ is the entropic injectivity describing the thermally induced charge injection into QD$_1$. As discussed above, a pure thermal bias does not generate a net subgap Andreev thermocurrent because of the intrinsic particle-hole symmetry of Andreev reflection. Nevertheless, the thermal bias induces a finite internal potential shift $\delta U(\theta)$ through the nonequilibrium screening response of QD$_1$. Since the anomalous contribution to the entropic injectivity vanishes identically, the resulting internal potential response originates entirely from the normal contribution, i.e., the first term in Eq.~\ref{eq:B11}.
\begin{figure*}[!htb]
\centering
\includegraphics
[width=0.42\hsize]
{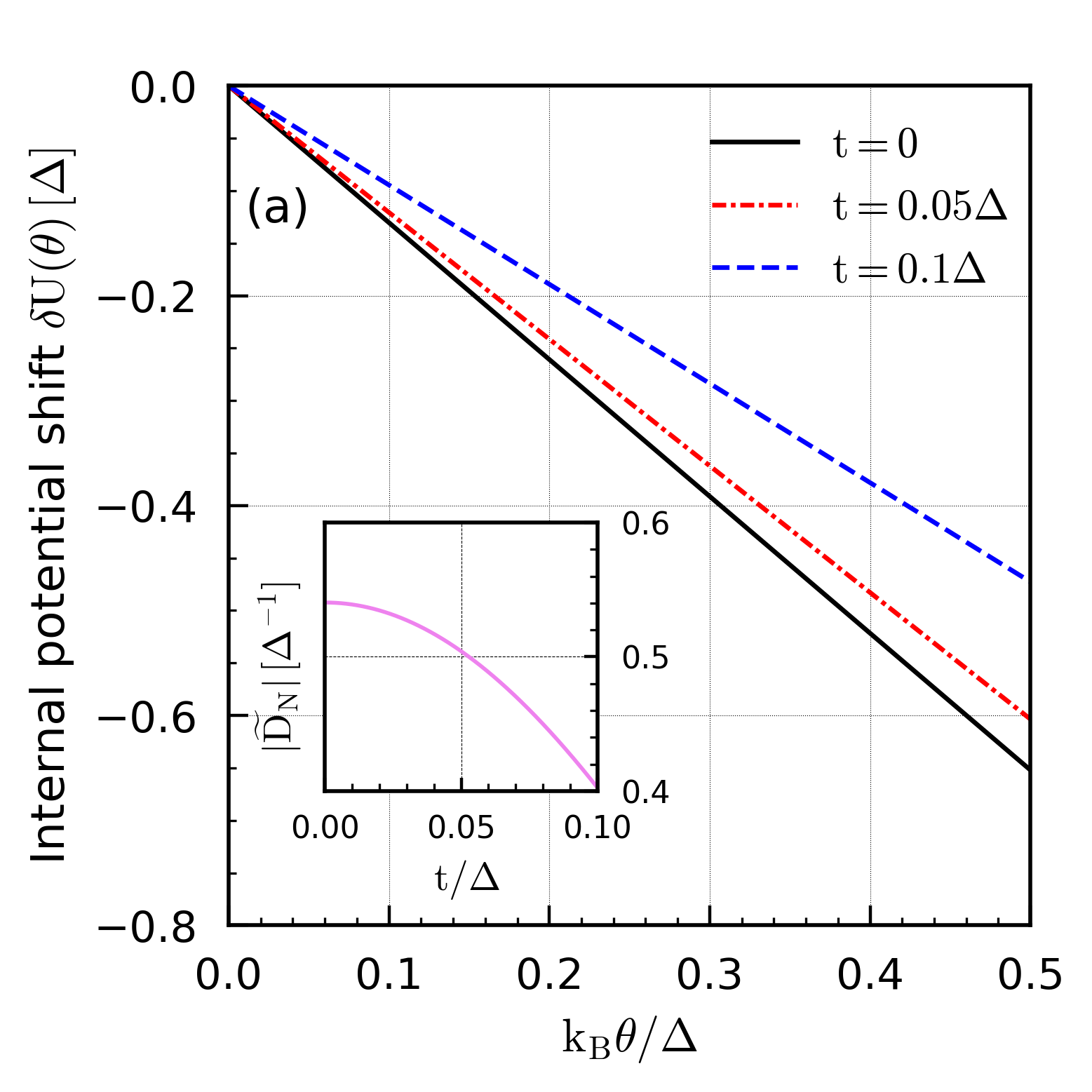}
\includegraphics
[width=0.42\hsize]
{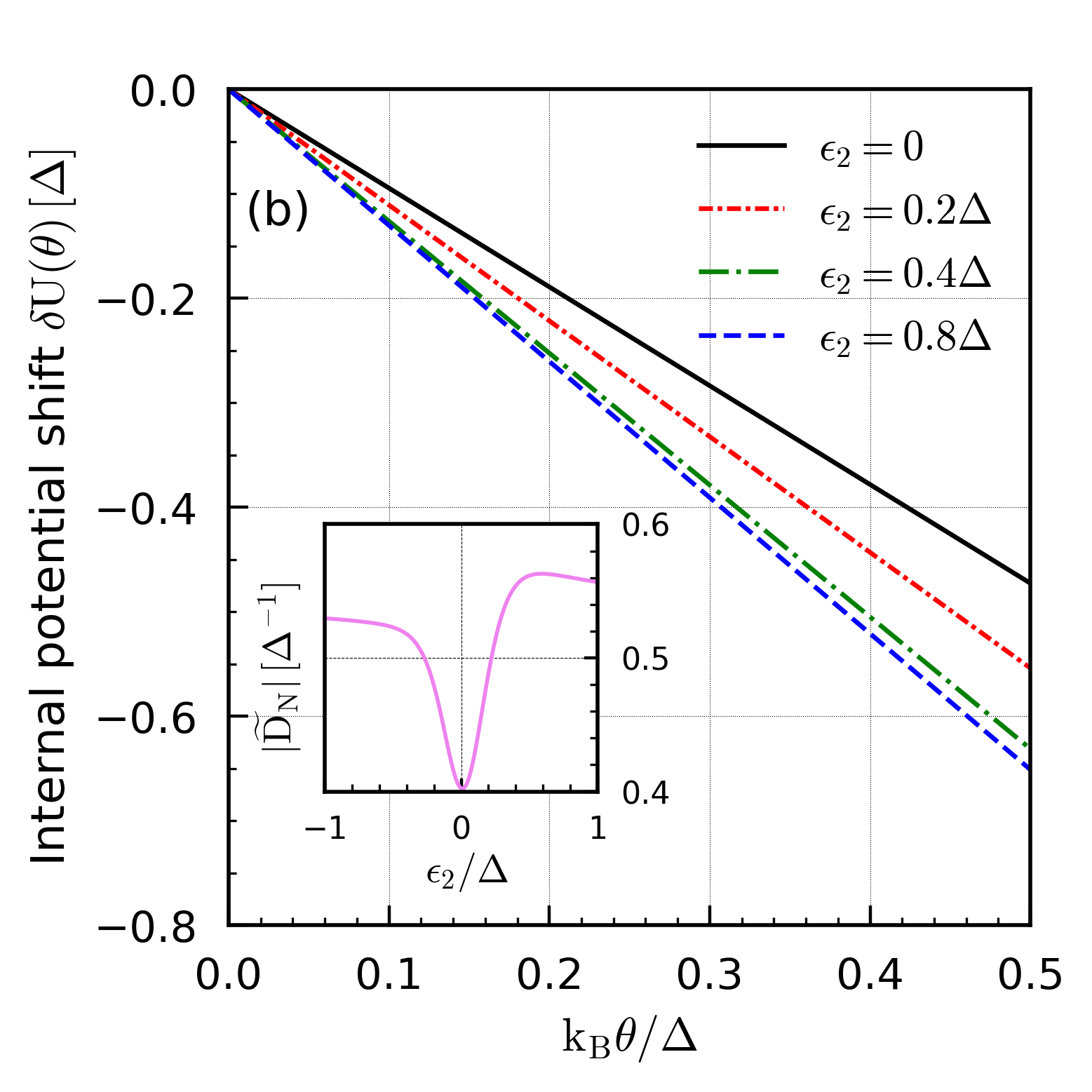}
\caption{(a) Internal potential shift, $\delta U(\theta)$, as a function of the thermal gradient $k_B\theta$ for different interdot couplings $t$, with $\epsilon_1=0.4\Delta$ and $\epsilon_2=0$. The inset shows the magnitude of the entropic injectivity, $\widetilde{D}_{N}$, as a function of the interdot coupling $t$. Since the anomalous contribution vanishes identically, $\widetilde{D}_{N}$ is entirely determined by the normal electronic
contribution. (b) $\delta U(\theta)$ as a function of $k_B\theta$ for different $\epsilon_2$, with $t=0.1\Delta$ and $\epsilon_1=0.4\Delta$. The inset shows the magnitude of the entropic injectivity, $\widetilde{D}_{N}$, as a function of $\epsilon_2$. The remaining parameters are $\Gamma_N=\Gamma_S=0.3\Delta$, and $k_B T=0.1\Delta$.}
\label{fig:5}
\end{figure*}\\
As shown in Fig.~\ref{fig:5}(a), for $\epsilon_1\neq 0$, the applied thermal bias induces a charge imbalance by probing the electron-hole asymmetry of the QD$_1$ spectrum. For the QD$_1$ level lying above the Fermi level, Eq.~\ref{eq:B11} yields $\widetilde{D}_N<0$, resulting in a negative thermal characteristic potential $z_N$. Consequently, since $\delta U\simeq z_N\theta$, the internal potential shifts toward negative values with increasing thermal bias, thereby screening the thermally induced charge imbalance. With $\tilde{\epsilon}_1=\epsilon_1-eU$, a negative $\delta U$ shifts the renormalized QD$_1$ level farther above the Fermi level. The magnitude of the internal potential shift is largest for $t=0$ and decreases with increasing interdot coupling $t$. As $t$ increases, the Fano-Andreev effect redistributes the electronic states of QD$_1$ around the Fermi level, reducing the thermally weighted imbalance between states above and below it. Since these states contribute with opposite signs to Eq.~\ref{eq:B11}, their stronger compensation reduces $|\widetilde{D}_N|$ [see inset of Fig.~\ref{fig:5}(a)] and consequently $|\delta U|$. Figure~\ref{fig:5}(b) shows that $|\delta U|$ increases as QD$_2$ level is detuned from the Fermi energy, following the corresponding increase in $|\widetilde{D}_N|$ shown in the inset. Near $\epsilon_2=0$, the Fano-Andreev effect suppresses $|\widetilde{D}_N|$, whereas increasing $|\epsilon_2|$ detunes the QD$_2$ level from the thermally active energy window, causing $|\widetilde{D}_N|$, and hence $|\delta U(\theta)|$, to increase and eventually approach their off-resonant values.
\begin{figure*}[!htb]
\centering
\includegraphics
[width=0.38\hsize]
{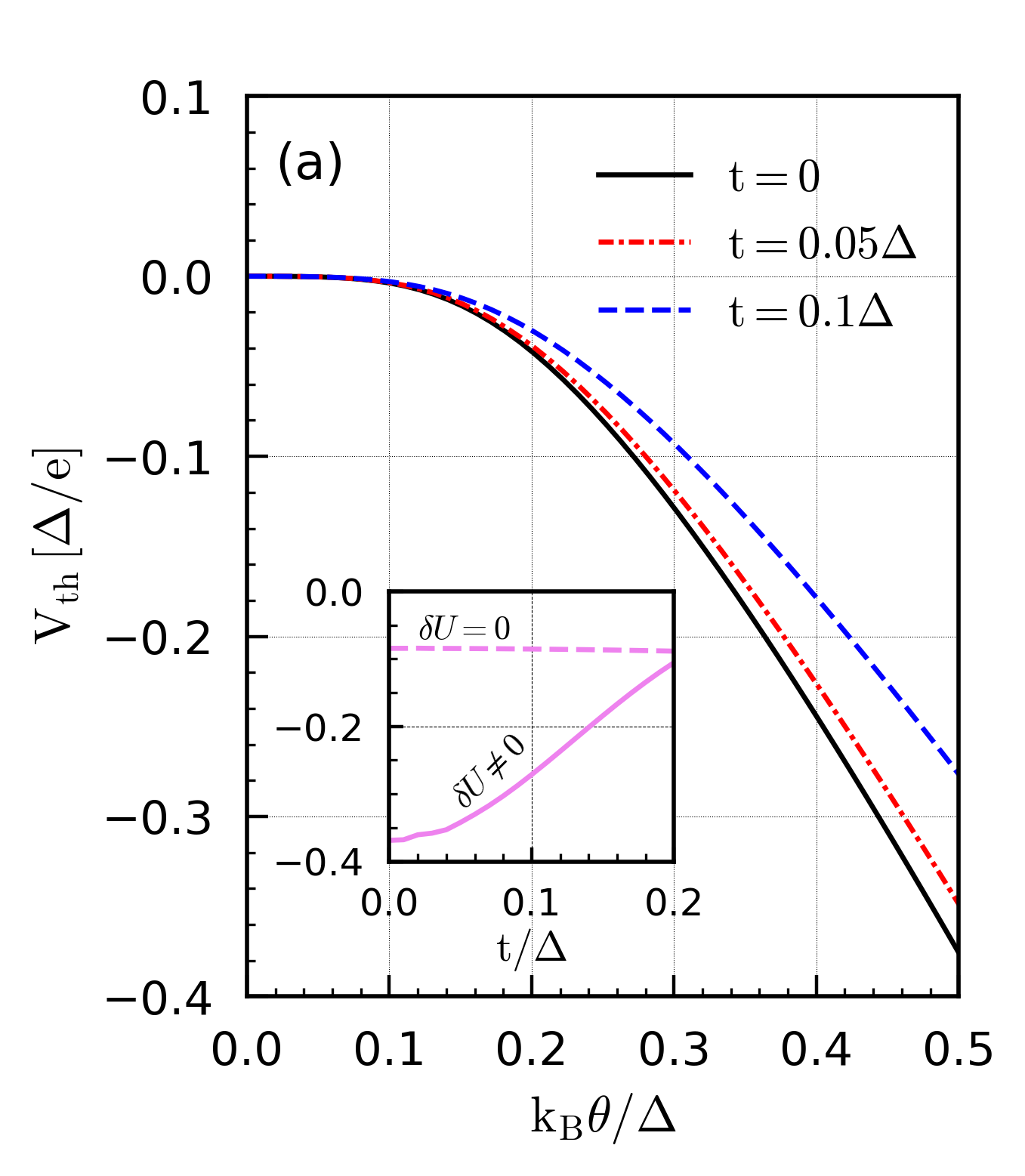}
\includegraphics
[width=0.38\hsize]
{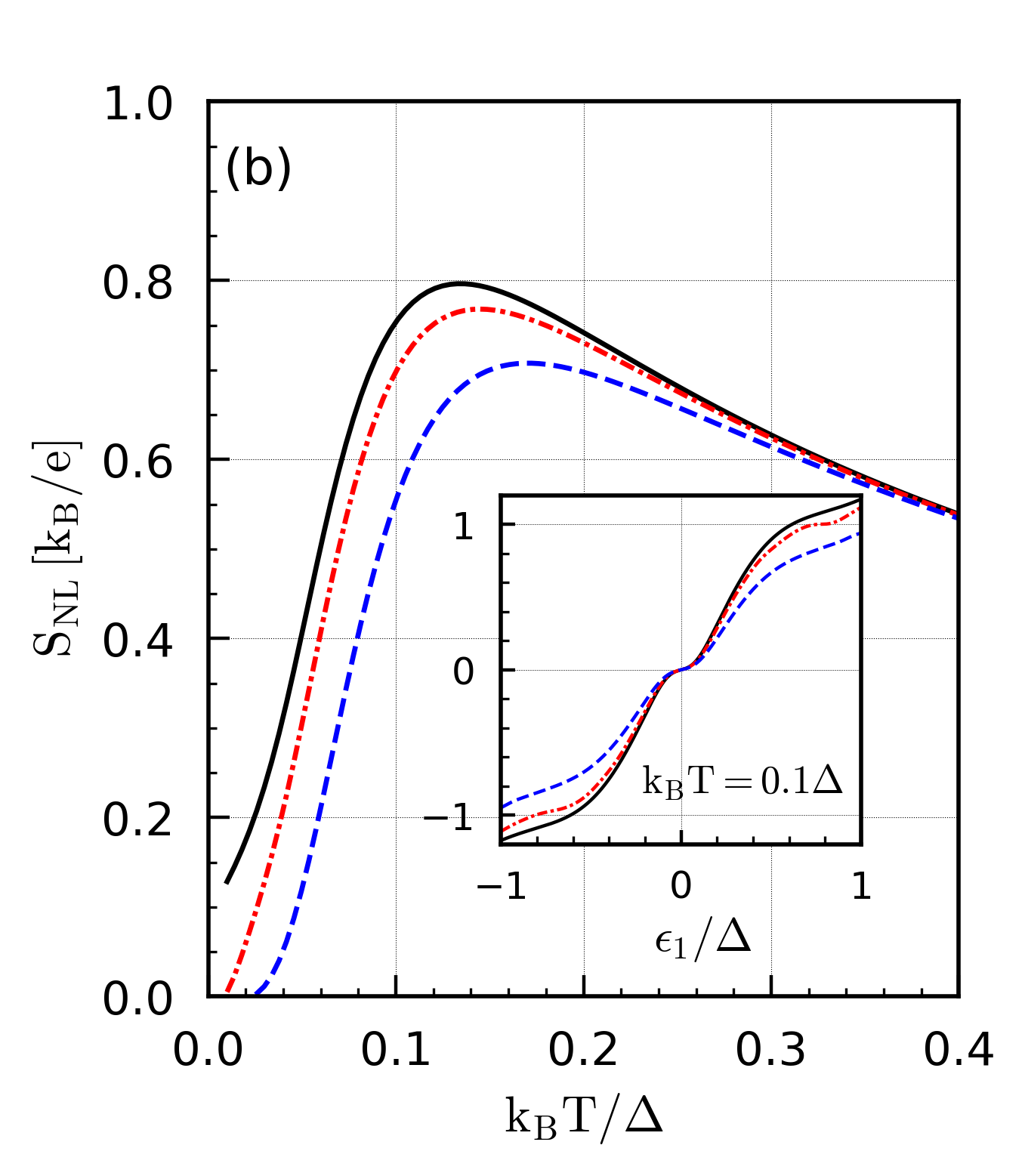}
\caption{(a) Thermovoltage, $V_{\mathrm{th}}$, as a function of the thermal bias $k_B\theta$ for different values of the interdot coupling $t$, with $\epsilon_1=0.4\Delta$ and $k_B T=0.1\Delta$. The inset compares $V_{\mathrm{th}}$ as a function of $t$ at a fixed thermal bias $k_B\theta=0.5\Delta$, with internal potential shift $\delta U\neq0$ (solid line) and with $\delta U=0$ (dashed line). (b) Nonlinear Seebeck coefficient or thermopower, $S_{NL}$, as a function of the background thermal energy $k_B T$ for different values of $t$, evaluated at a fixed thermal bias $k_B\theta=0.5\Delta$, with $\epsilon_1=0.4\Delta$. The inset shows $S_{NL}$ as a function of the $\epsilon_1$ for different values of $t$ at $k_B\theta=0.5\Delta$ and $k_B T=0.1\Delta$. The remaining parameters are $\epsilon_2=0$ and $\Gamma_N=\Gamma_S=0.3\Delta$.}
\label{fig:6}
\end{figure*}\\
The behavior of the thermovoltage $V_{\mathrm{th}}$ and nonlinear Seebeck coefficient (thermopower) $S_{NL}$ is closely related to the thermal screening response discussed in Fig.~\ref{fig:6}. For $\epsilon_1>0$, the thermally induced internal potential shifts the effective QD$_1$ level farther above the Fermi energy and modifies the quasiparticle spectrum. Since $V_{\mathrm{th}}$ and $S_{NL}$ arise from quasiparticle transport above the superconducting gap ($|\omega|>\Delta$), this spectral renormalization influences the thermoelectric response. As shown in Fig.~\ref{fig:6}(a), the thermovoltage $V_{\mathrm{th}}$ remains very small at low thermal biases and increases significantly only at larger $\theta$. A thermal bias alone does not generate an Andreev thermocurrent because of particle-hole symmetry, whereas a finite thermovoltage arises from quasiparticle transport as the thermal bias increases. The magnitude of $V_{\mathrm{th}}$ decreases with increasing interdot coupling $t$. In contrast, when the internal potential shift is neglected ($\delta U=0$), $V_{\mathrm{th}}$ becomes nearly independent of $t$, as shown in the inset. This demonstrates that the $t$ dependence of $V_{\mathrm{th}}$ originates predominantly from the thermal screening response. As $t$ increases, $|\delta U(\theta)|$ decreases, weakening the screening-induced renormalization of the QD$_1$ level and thereby reducing the magnitude of the compensating voltage required to satisfy the open-circuit condition $I_C(V_{\mathrm{th}},\theta)=0$.\\
Figure~\ref{fig:6}(b) shows the nonlinear Seebeck coefficient, $S_{NL}$, calculated within the first-order characteristic-potential approximation as a function of the background thermal energy $k_B T$. For all values of $t$, $S_{NL}$ exhibits a nonmonotonic temperature dependence, initially increasing as quasiparticle transport becomes thermally activated, reaching a maximum, and subsequently decreasing at higher temperatures due to thermal broadening. The magnitude of $S_{NL}$ decreases with increasing interdot coupling $t$, consistent with the reduced thermovoltage and weaker screening-induced renormalization of the QD$_1$ level. The inset shows $S_{NL}$ as a function of $\epsilon_1$ at fixed $k_B T=0.1\Delta$ and $k_B\theta=0.5\Delta$. The thermopower vanishes at the particle-hole symmetric point $\epsilon_1=0$ and increases in magnitude as the QD$_1$ level is tuned away from the Fermi energy, reflecting the increasing electron-hole asymmetry of the quasiparticle spectrum. Overall, increasing $t$ reduces the magnitude of $S_{NL}$ over the entire range of $\epsilon_1$, consistent with the reduced screening-induced level renormalization.\\

\paragraph*{\textbf{Simultaneous voltage and thermal bias (cross thermoelectric coupling):}} We now consider the nonequilibrium regime in which small but finite voltage and temperature biases are applied simultaneously. As discussed above, in the weakly nonlinear regime the leading-order internal potential shift is given by $\delta U(V,\theta)=u_NV+z_N\theta$. Unlike the pure voltage or pure thermal cases, the simultaneous electrical and thermal driving gives rise to a finite cross thermoelectric response. In the following, we investigate how nonlinear screening and the Fano-Andreev effect govern the low-bias Andreev transport and the resulting cross thermoelectric response.
\begin{figure*}[!htb]
\centering
\includegraphics[width=0.32\hsize]{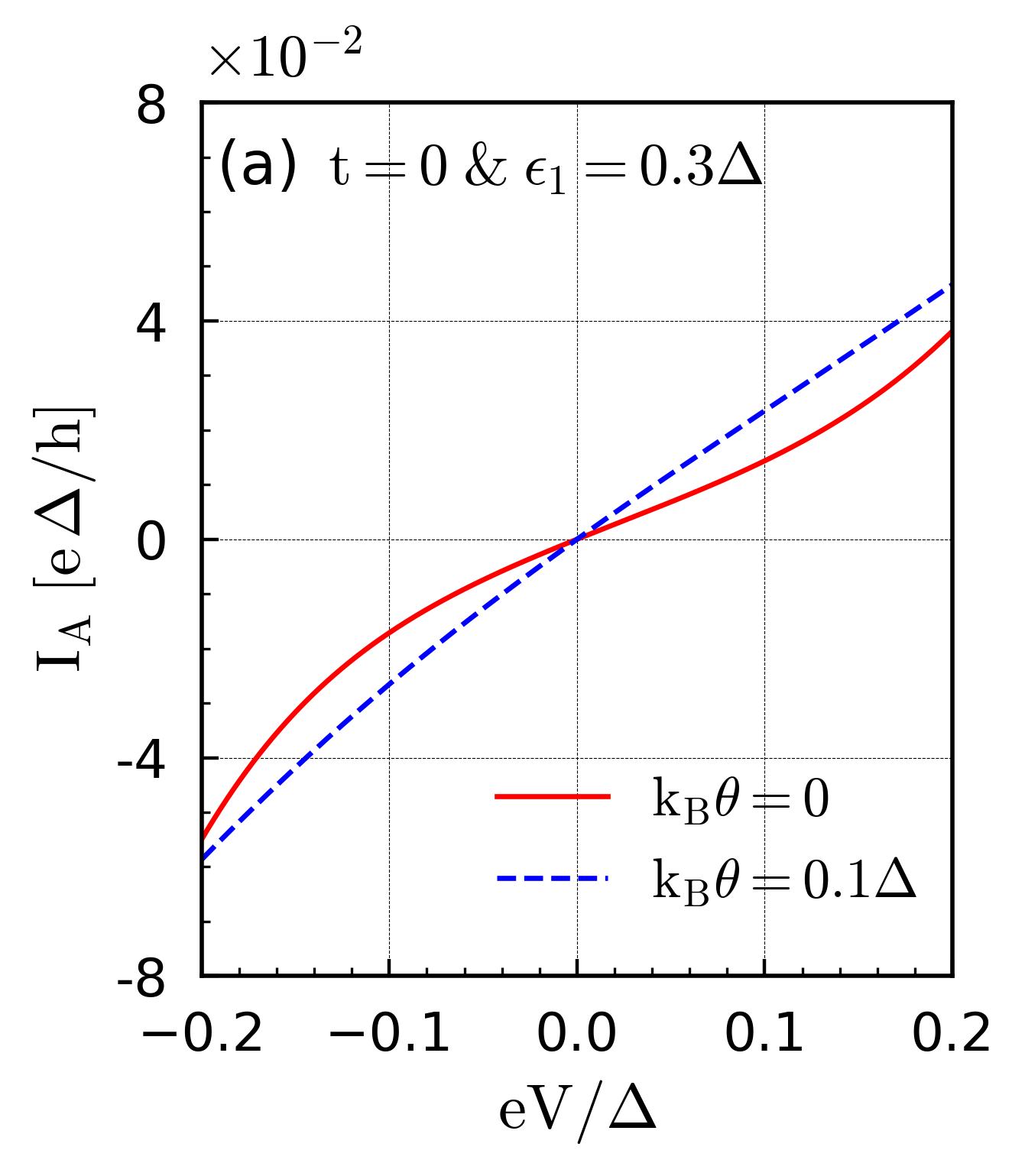}\hspace{0.01\hsize}
\includegraphics[width=0.32\hsize]{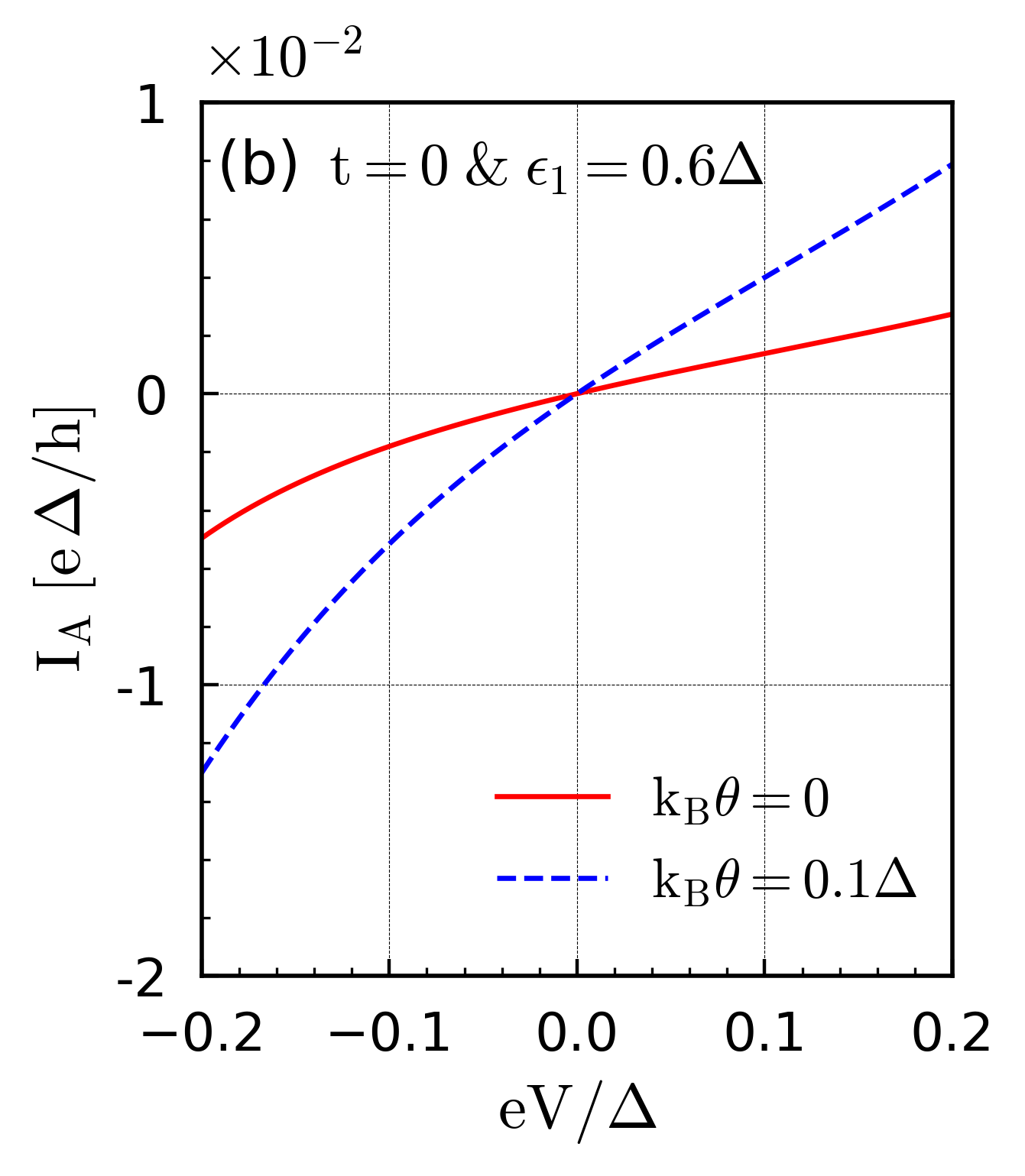}\\[1ex]
\includegraphics[width=0.32\hsize]{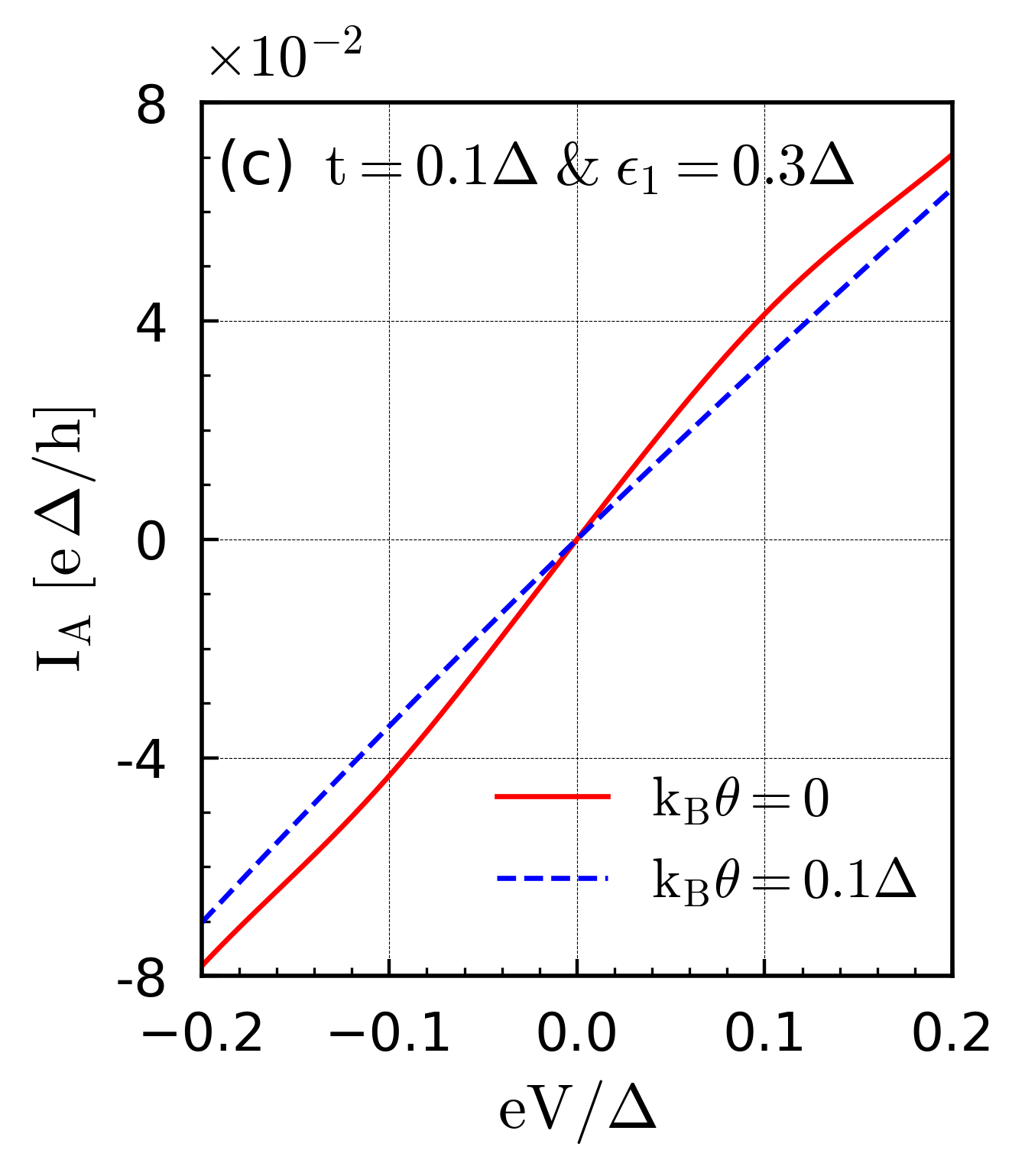}\hspace{0.01\hsize}
\includegraphics[width=0.32\hsize]{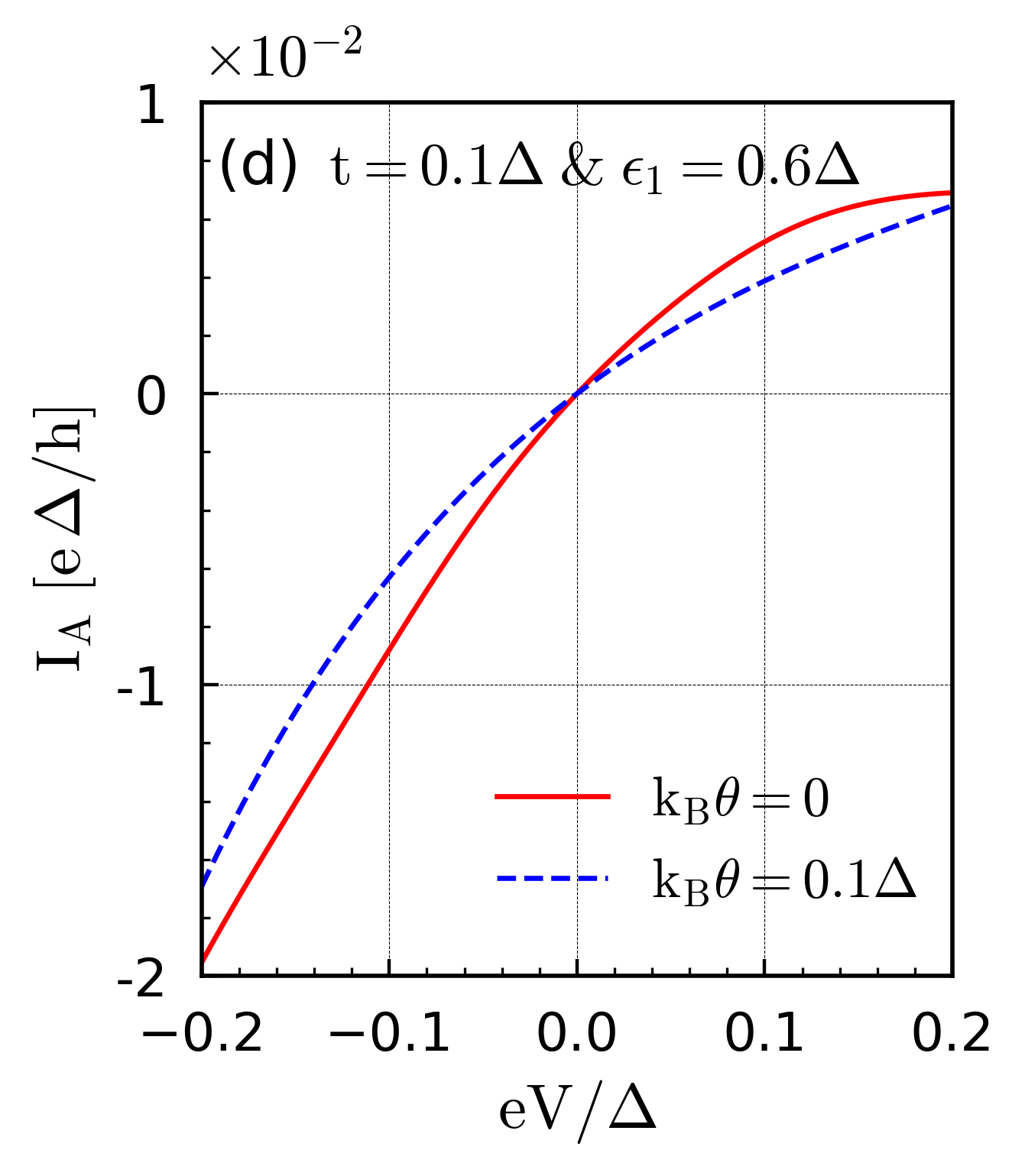}
\caption{Low-bias Andreev current-voltage ($I_A-V$) characteristics for different QD$_1$ energy levels $\epsilon_1$ and interdot couplings $t$. Panels (a)-(b) correspond to the single-dot configuration ($t=0$) with $\epsilon_1=0.3\Delta$, and $0.6\Delta$, respectively. Panels (c)-(d) show the corresponding results for the coupled double-dot configuration with $t=0.1\Delta$ and $\epsilon_2=0.1\Delta$, for $\epsilon_1=0.3\Delta$, and $0.6\Delta$, respectively. In all panels, the results are shown for temperature biases $k_B\theta=0$ and $0.1\Delta$. The remaining parameters are $\Gamma_N=0.1\Delta$, $\Gamma_S=0.5\Delta$, and $k_BT=0.05\Delta$.}
\label{fig:7}
\end{figure*}\\
Figure~\ref{fig:7} shows the low-bias Andreev transport characteristics for $|eV|<0.2\Delta$ and $k_B\theta<0.1\Delta$ at background thermal energy $k_BT=0.05\Delta$, where transport is dominated by coherent Andreev reflection and the contribution from quasiparticle tunneling is negligible. The response of the Andreev current to a thermal bias depends sensitively on the interdot coupling $t$. For the single-dot configuration ($t=0$) [Figs.~\ref{fig:7}(a) and \ref{fig:7}(b)], increasing the temperature bias enhances the low-bias Andreev current for both $\epsilon_1=0.3\Delta$ and $\epsilon_1=0.6\Delta$. In contrast, for the coupled double-dot configuration ($t=0.1\Delta$ and $\epsilon_2=0.1\Delta$) [Figs.~\ref{fig:7}(c) and \ref{fig:7}(d)], the Andreev current is progressively suppressed throughout the low-bias regime as the temperature bias increases. The origin of this contrasting behavior can be understood from the cross-thermoelectric coefficient $M_1$, defined in Eq.~\ref{eq:24}, which characterizes the leading-order coupling between the voltage and thermal biases in the weakly nonlinear regime. The sign of $M_1$ determines whether the thermal bias enhances or suppresses the Andreev current: a positive (negative) $M_1$ leads to an enhancement (suppression) of the low-bias $I_A-V$ characteristics with increasing thermal bias. Physically, $M_1$ is determined by the competition between the two terms in the integrand of Eq.~\ref{eq:24}. The first term represents the screening contribution through the characteristic potential $z_N$, which accounts for the temperature-induced charge redistribution and the resulting modification of the transmission by the internal electrostatic potential. The second term represents the intrinsic transport contribution arising from the energy dependence of the transmission and remains finite even in the absence of screening.
\begin{figure*}[!htb]
\centering
\includegraphics
[width=0.63\hsize]
{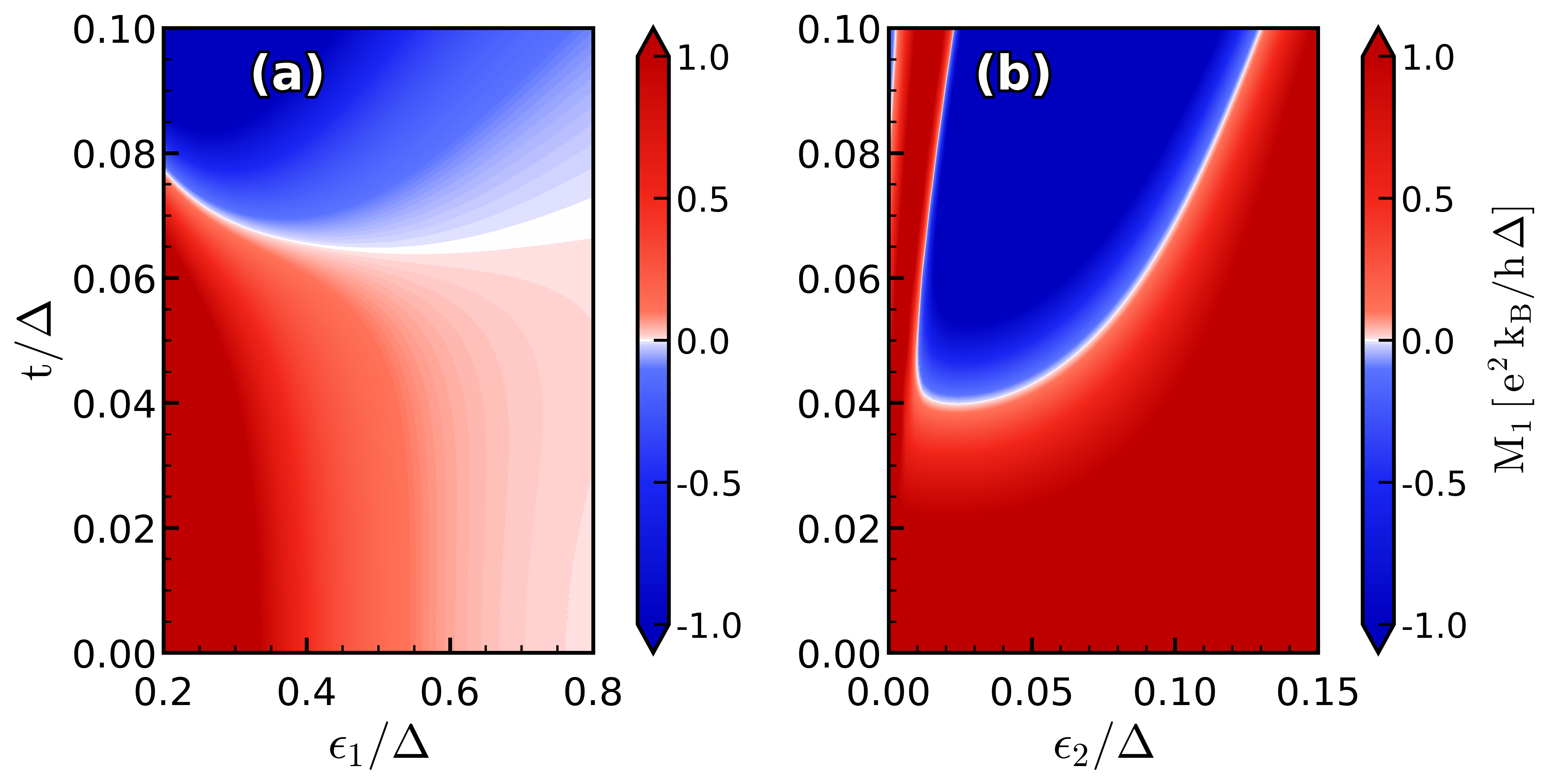}
\includegraphics
[width=0.63\hsize]
{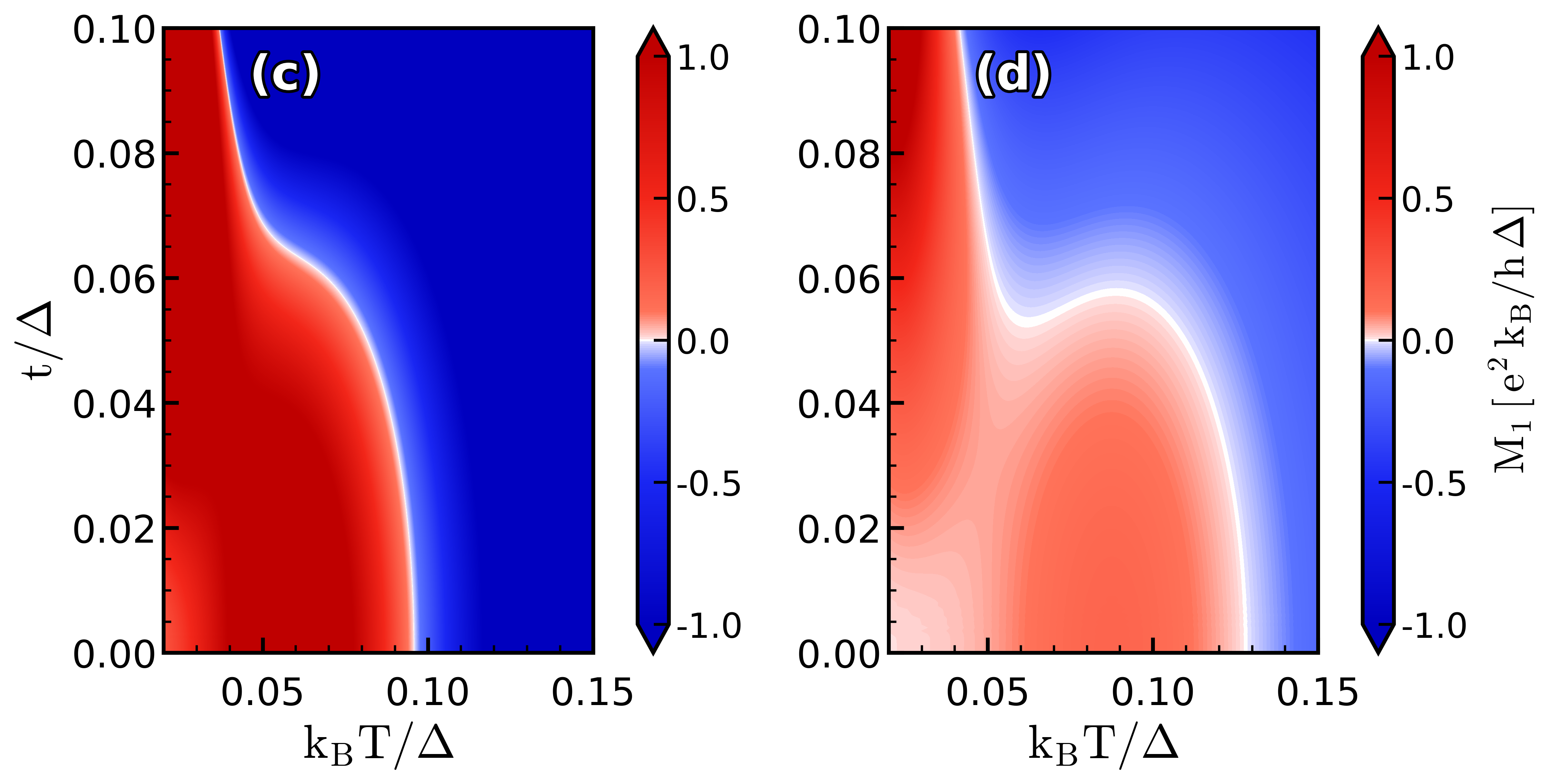}
\caption{Color maps of the cross thermoelectric coupling coefficient $M_1$. Panel (a) shows $M_1$ as a function of the QD$_1$ energy level $\epsilon_1$ and the interdot coupling $t$ for $\epsilon_2=0.1\Delta$ at the background thermal energy $k_BT=0.05\Delta$. Panel (b) shows $M_1$ as a function of QD$_2$ energy level $\epsilon_2$ and $t$ for $\epsilon_1=0.3\Delta$ at $k_BT=0.05\Delta$. Panels (c) and (d) show $M_1$ as a function of the background thermal energy $k_BT$ and the interdot coupling $t$ for $\epsilon_2=0.1\Delta$ with $\epsilon_1=0.3\Delta$ and $0.6\Delta$, respectively. The remaining parameters are $\Gamma_N=0.1\Delta$ and $\Gamma_S=0.5\Delta$.}
\label{fig:8}
\end{figure*}\\
Figure~\ref{fig:8}(a) shows the variation of the cross thermoelectric coefficient $M_1$ as a function of the QD$_1$ energy level $\epsilon_1$ and the interdot coupling $t$ at $k_BT=0.05\Delta$. Here, we focus on the region $\epsilon_1\geq 0.2\Delta$, where the sign reversal of $M_1$ is governed by the competition between the energy-dependent contribution and screening contribution. For weak interdot coupling, the positive energy-dependent contribution dominates, resulting in $M_1>0$. As the interdot coupling increases, the Fano-Andreev effect modifies the energy dependence of the Andreev transmission within the thermally active window, reducing its net thermally weighted slope and thereby suppressing the positive energy-dependent contribution to $M_1$. Consequently, the negative screening contribution becomes dominant, driving the sign reversal of $M_1$ and producing a pronounced region with $M_1<0$. Figure~\ref{fig:8}(b) shows $M_1$ as a function of QD$_2$ level $\epsilon_2$ and the interdot coupling $t$ at $\epsilon_1=0.3\Delta$ and $k_BT=0.05\Delta$. Near $\epsilon_2\approx 0$, the positive energy-dependent contribution remains larger than the negative screening contribution, resulting in $M_1>0$. With increasing $\epsilon_2$, the Fano-Andreev effect modifies the thermally weighted energy dependence of the Andreev transmission and suppresses the positive contribution, allowing the negative screening contribution to dominate and driving $M_1<0$. At the higher end of the considered $\epsilon_2$ range, the influence of the Fano-Andreev features weakens, altering the balance between the energy-dependent and screening contributions and leading to the recovery of positive $M_1$.\\
Figure~\ref{fig:8}(c) shows the variation of the cross thermoelectric coefficient $M_1$ as a function of the background thermal energy $k_BT$ and $t$ at $\epsilon_1=0.3\Delta$ and $\epsilon_2=0.1\Delta$. At low temperatures and weak interdot coupling, the positive energy-dependent contribution exceeds the negative screening contribution, resulting in $M_1>0$. As the background temperature and interdot coupling increase, the balance between these contributions shifts toward the negative screening term, which eventually dominates and drives the sign reversal of $M_1$. Consequently, the boundary between the positive and negative regions shifts toward lower temperatures with increasing interdot coupling. Figure~\ref{fig:8}(d) shows the corresponding results for $\epsilon_1=0.6\Delta$. Compared with Fig.~\ref{fig:8}(c), the positive-$M_1$ region extends over a broader range of background temperatures and interdot couplings. This behavior reflects the modified energy dependence of the Andreev transmission at $\epsilon_1=0.6\Delta$, for which the positive energy-dependent contribution remains sufficient to compensate the negative screening contribution over a wider parameter range.

\section{Conclusion}
We have investigated weakly nonlinear electrical and thermoelectric transport through a T-shaped double quantum dot coupled to normal and superconducting leads, with nonlinear screening effects incorporated within the characteristic-potential approach. The side-coupled quantum dot gives rise to the Fano-Andreev effect, evolving from Fano-type interference in the weak-coupling regime to hybridized Andreev-Fano states in the molecular regime, and thereby strongly influencing the nonequilibrium transport response.\\
Under pure voltage driving, the internal screening potential produces a bias-dependent renormalization of lead-coupled central quantum dot energy level. At low temperatures, low-bias Andreev transport is governed primarily by the Fano-Andreev effect, with nonlinear screening introducing only a weak bias asymmetry, whereas at higher temperatures the influence of nonlinear screening becomes more pronounced, resulting in a strong bias asymmetry in the Andreev differential conductance. Under pure thermal driving, increasing the interdot coupling weakens the thermal screening response through the Fano-Andreev-induced redistribution of electronic states, thereby reducing the screening-induced renormalization of the central dot energy level and suppressing the above-gap quasiparticle thermoelectric response, which in turn reduces the magnitudes of the thermovoltage and nonlinear Seebeck coefficient.\\
Furthermore, simultaneous voltage and thermal biases generate a finite subgap cross thermoelectric response beyond linear response. Its sign is governed by the competition between the energy-dependent transport contribution and the nonlinear screening contribution. The Fano-Andreev effect modifies the thermally weighted energy dependence of the Andreev transmission, thereby altering the balance between these two contributions and leading to sign reversals of the cross thermoelectric coefficient. Consequently, a thermal bias can either enhance or suppress the low-bias Andreev current, while the interdot coupling can further tune this response through the Fano-Andreev effect, providing an effective mechanism for controlling nonlinear subgap transport.
\label{sec:4}

\appendix
\section{Green's Functions on the Center Quantum Dot}
\label{App:A}
In this Appendix, we derive the effective Nambu retarded Green’s function for the lead-coupled QD$_1$. In the present configuration, only QD$_1$ is directly coupled to the leads, while QD$_2$ is coupled to the QD$_1$ via interdot tunneling $t$.\\
We first construct the full Green’s function in Nambu space and subsequently integrate out QD$_2$ to obtain effective descriptions for QD$_1$. We define the Nambu spinor for the double quantum dot system as
\begin{equation}
\Psi = (d_{1\uparrow},\, d_{2\uparrow},\, d^\dagger_{1\downarrow},\, d^\dagger_{2\downarrow})^{T}.
\label{eq:A1}
\end{equation}
In this basis, the retarded Green's function matrix in Nambu space is defined using the Zubarev double-time Green's function formalism as
\begin{equation}
\mathbf{G}^{r}(\omega) = \langle\!\langle \Psi \, ; \, \Psi^\dagger \rangle\!\rangle_\omega,
\label{eq:A2}
\end{equation}
which can be written explicitly as
\begin{equation}
\renewcommand{\arraystretch}{1.4}
\mathbf{G}^{r}(\omega) =
%
%
%
\begin{pmatrix}
\langle\!\langle d_{1\uparrow}; d_{1\uparrow}^\dagger \rangle\!\rangle_\omega \quad &
\langle\!\langle d_{1\uparrow}; d_{2\uparrow}^\dagger \rangle\!\rangle_\omega \quad &
\langle\!\langle d_{1\uparrow}; d_{1\downarrow} \rangle\!\rangle_\omega \quad &
\langle\!\langle d_{1\uparrow}; d_{2\downarrow} \rangle\!\rangle_\omega \\

\langle\!\langle d_{2\uparrow}; d_{1\uparrow}^\dagger \rangle\!\rangle_\omega \quad &
\langle\!\langle d_{2\uparrow}; d_{2\uparrow}^\dagger \rangle\!\rangle_\omega \quad &
\langle\!\langle d_{2\uparrow}; d_{1\downarrow} \rangle\!\rangle_\omega \quad &
\langle\!\langle d_{2\uparrow}; d_{2\downarrow} \rangle\!\rangle_\omega \\

\langle\!\langle d_{1\downarrow}^\dagger; d_{1\uparrow}^\dagger \rangle\!\rangle_\omega \quad &
\langle\!\langle d_{1\downarrow}^\dagger; d_{2\uparrow}^\dagger \rangle\!\rangle_\omega \quad &
\langle\!\langle d_{1\downarrow}^\dagger; d_{1\downarrow} \rangle\!\rangle_\omega \quad &
\langle\!\langle d_{1\downarrow}^\dagger; d_{2\downarrow} \rangle\!\rangle_\omega \\

\langle\!\langle d_{2\downarrow}^\dagger; d_{1\uparrow}^\dagger \rangle\!\rangle_\omega \quad &
\langle\!\langle d_{2\downarrow}^\dagger; d_{2\uparrow}^\dagger \rangle\!\rangle_\omega \quad &
\langle\!\langle d_{2\downarrow}^\dagger; d_{1\downarrow} \rangle\!\rangle_\omega \quad &
\langle\!\langle d_{2\downarrow}^\dagger; d_{2\downarrow} \rangle\!\rangle_\omega
\end{pmatrix}.
\label{eq:A3}
\end{equation}
The upper-left $2 \times 2$ block corresponds to the normal (electron) Green's functions, while the upper-right block describes anomalous (pairing) correlations arising from superconducting proximity effects. The lower blocks correspond to the hole sector and are related to the upper blocks by particle-hole symmetry. The Green's function in Zubarev notation is defined as
\begin{equation}
\langle\!\langle A ; B \rangle\!\rangle_\omega
=
-i \int_{0}^{\infty} dt \, e^{i\omega t} \langle \{ A(t), B(0) \} \rangle,
\label{eq:A4}
\end{equation}
where $\{\cdots\}$ denotes the anticommutator and $\langle\cdots\rangle$ denotes the thermal average over the equilibrium many-body states. By using the equation of motion method and Dyson equation, we obtain the retarded Green's function matrix of the system as
\begin{equation}
\mathbf{G}^r(\omega)
=
\Big[\mathbf{g}^r(\omega)^{-1}
-
\mathbf{\Sigma}_N^r
-
\mathbf{\Sigma}_S^r(\omega)\Big]^{-1}.
\label{eq:A5}
\end{equation}
The inverse of retarded Green's function of the T-shaped DQD without coupling to the leads is
\begin{equation}
\mathbf{g}^r(\omega)^{-1} =
\begin{pmatrix}
\omega-\tilde{\epsilon}_1 & t & 0 & 0 \\[6pt]
t & \omega-\epsilon_2 & 0 & 0 \\[6pt]
0 & 0 & \omega+\tilde{\epsilon}_1 & -t \\[6pt]
0 & 0 & -t & \omega+\epsilon_2
\end{pmatrix},
\label{eq:A6}
\end{equation}
with $\omega \to \omega + i0^+$. Since only QD$_1$ is coupled to the leads, the total retarded self-energies under the wide-bandwidth approximation are given by
\begin{equation}
\boldsymbol{\Sigma}^r(\omega)
= \boldsymbol{\Sigma}^r_N + \boldsymbol{\Sigma}^r_S(\omega)
= -\frac{i}{2}
\setlength{\arraycolsep}{10pt}
\renewcommand{\arraystretch}{1.5}
\begin{pmatrix}
\Gamma_N + \Gamma_S \beta(\omega) & 0 & -\Gamma_S\,\dfrac{\Delta}{|\omega|}\,\beta(\omega) & 0 \\
0 & 0 & 0 & 0 \\
-\Gamma_S\,\dfrac{\Delta}{|\omega|}\,\beta(\omega) & 0 &
\Gamma_N + \Gamma_S \beta(\omega) & 0 \\
0 & 0 & 0 & 0
\end{pmatrix}.
\label{eq:A7}
\end{equation}
To obtain the Green's function for QD$_1$, we integrate out QD$_2$ by performing a block inversion of Eq.~\ref{eq:A5}. The retarded Green’s function of the isolated QD$_2$ is given by
\begin{equation}
\mathbf{g}_2^r(\omega)=
\begin{pmatrix}
\dfrac{1}{\omega-\epsilon_2} & 0 \\[6pt]
0 & \dfrac{1}{\omega+\epsilon_2}
\end{pmatrix},
\label{eq:A8}
\end{equation}
which yields the effective self-energy contribution
\begin{equation}
\mathbf{\Sigma}_{\mathrm{QD_2}}(\omega)
=
t^2\,\mathbf{g}_2^r(\omega)
=
\begin{pmatrix}
\dfrac{t^2}{\omega-\epsilon_2} & 0 \\[6pt]
0 & \dfrac{t^2}{\omega+\epsilon_2}
\end{pmatrix}.
\label{eq:A9}
\end{equation}
The effective $2\times2$ Nambu Green’s function of the QD$_1$ is then obtained as
\begin{equation}
\mathbf{G^r_{QD_1}}(\omega)
=
\begin{pmatrix}
G_{11}^r(\omega) & G_{13}^r(\omega) \\[6pt]
G_{31}^r(\omega) & G_{33}^r(\omega)
\end{pmatrix}
=
\left[
\begin{pmatrix}
\omega - \tilde{\epsilon}_1 & 0 \\[6pt]
0 & \omega + \tilde{\epsilon}_1
\end{pmatrix}
-
\mathbf{\Sigma}_{\mathrm{QD_2}}(\omega)
-
\mathbf{\Sigma}^r_N
-
\mathbf{\Sigma}^r_S(\omega)
\right]^{-1}.
\label{eq:A10}
\end{equation}
Substituting all contributions, we obtain
\begin{equation}
\mathbf{G^r_{QD_1}}(\omega)
=
\begin{pmatrix}
\omega - \tilde{\epsilon}_1 - \dfrac{t^2}{\omega-\epsilon_2}  + \dfrac{i}{2}\Gamma_N + \dfrac{i}{2}\Gamma_S\,\beta(\omega) 
&
-\dfrac{i}{2}\Gamma_S\,\dfrac{\Delta}{|\omega|}\,\beta(\omega)
\\[10pt]
-\dfrac{i}{2}\Gamma_S\,\dfrac{\Delta}{|\omega|}\,\beta(\omega)
&
\omega + \tilde{\epsilon}_1 - \dfrac{t^2}{\omega+\epsilon_2} + \dfrac{i}{2}\Gamma_N + \dfrac{i}{2}\Gamma_S\,\beta(\omega)
\end{pmatrix}^{-1}.
\label{eq:A11}
\end{equation}
The matrix inversion yields explicit expressions for the Green’s function components on QD$_1$. The normal electronic component is given by
\begin{equation}
G_{11}^r(\omega)
=
\frac{
\omega + \tilde{\epsilon}_1 
- \dfrac{t^2}{\omega+\epsilon_2}
+ \dfrac{i}{2}\Gamma_N 
+ \dfrac{i}{2}\Gamma_S\,\beta(\omega) 
}{
\Big({\omega - \tilde{\epsilon}_1 
- \dfrac{t^2}{\omega-\epsilon_2}
+ \dfrac{i}{2}\Gamma_N 
+ \dfrac{i}{2}\Gamma_S\,\beta(\omega)}\Big)\Big({\omega + \tilde{\epsilon}_1 
- \dfrac{t^2}{\omega+\epsilon_2}
+ \dfrac{i}{2}\Gamma_N 
+ \dfrac{i}{2}\Gamma_S\,\beta(\omega)}\Big)
-\left(\dfrac{i}{2}\Gamma_S\,\dfrac{\Delta}{|\omega|}\,\beta(\omega)\right)^2},
\label{eq:A12}
\end{equation}
while the anomalous (particle-hole conversion) component reads
\begin{equation}
G_{13}^r(\omega)
=
\frac{
\dfrac{i}{2}\Gamma_S\,\dfrac{\Delta}{|\omega|}\,\beta(\omega)
}{
\Big({\omega - \tilde{\epsilon}_1 
- \dfrac{t^2}{\omega-\epsilon_2}
+ \dfrac{i}{2}\Gamma_N 
+ \dfrac{i}{2}\Gamma_S\,\beta(\omega)}\Big)\Big({\omega + \tilde{\epsilon}_1 
- \dfrac{t^2}{\omega+\epsilon_2}
+ \dfrac{i}{2}\Gamma_N 
+ \dfrac{i}{2}\Gamma_S\,\beta(\omega)}\Big)
-\left(\dfrac{i}{2}\Gamma_S\,\dfrac{\Delta}{|\omega|}\,\beta(\omega)\right)^2}.
\label{eq:A13}
\end{equation}
The remaining components follow from particle-hole symmetry and the symmetry of the Nambu matrix,
\begin{equation}
G_{33}^{r}(\omega)
=
-\left[G_{11}^{r}(-\omega)\right]^{*},
\qquad
G_{31}^{r}(\omega)
=
G_{13}^{r}(\omega),
\label{eq:A14}
\end{equation}
while the advanced components satisfy
\begin{equation}
G_{11}^{a}(\omega)
=
\left[G_{11}^{r}(\omega)\right]^{*},
\qquad
G_{13}^{a}(\omega)
=
\left[G_{13}^{r}(\omega)\right]^{*}.
\label{eq:A15}
\end{equation}
\section{Nonlinear Transport and Screening Effects}
\label{App:B}
Nonlinear transport in nanoscale systems is influenced by weak screening effects and nonequilibrium charge redistribution. A finite voltage and thermal bias leads to charge accumulation, which modifies transport via electrostatic feedback and shifts the dot energy levels~\cite{Buttiker1993,Michalek2015}. In the present T-shaped configuration, electron-electron interactions are included only on QD$_1$, which is directly coupled to the leads, while interactions on QD$_2$ are neglected. This approximation is justified because the nonequilibrium charge on QD$_2$ is induced indirectly via interdot tunneling and, in the weak interdot coupling limit, is typically of order $t^2$, making its contribution negligible.\\
Accordingly, the interaction-induced potential is included through the renormalization of the QD$_1$ level as
$\tilde{\epsilon}_1 \rightarrow \epsilon_1 - e\,U(V,\theta)$, where $U(V,\theta)$ denotes the voltage bias- and temperature-induced internal potential. Under finite voltage and thermal biases, the potential deviates from equilibrium and, for weak driving, can be expanded to leading order as~\cite{Hwang2015,Meair2013,Sanchez2013}
\begin{equation}
\delta U = U - U_{\mathrm{eq}} =
\sum_{\alpha\in N,S} \left( u_{\alpha}\,V_\alpha + z_{\alpha}\,\theta_\alpha \right),
\label{eq:B1}
\end{equation}
where $u_{\alpha} = \left(\partial U/\partial V_\alpha\right)_{\mathrm{eq}}$ and $z_{\alpha} = \left(\partial U/\partial \theta_\alpha\right)_{\mathrm{eq}}$ are the characteristic potentials describing the electrical and thermal response of the QD$_1$. At equilibrium, the constant potential $U_{\mathrm{eq}}$ is absorbed into the definition of the renormalized level and is therefore not treated explicitly.\\
We consider that the normal lead is biased, such that $\mu_N = eV$ and $T_N = T + \theta$, while the superconducting lead remains at equilibrium with $\mu_S = 0$ and $T_S = T$. Under these conditions, the nonequilibrium screening potential is driven solely by the normal lead and reduces to
\begin{equation}
\delta U(V,\theta) = u_N\,V + z_N\,\theta.
\label{eq:B2}
\end{equation}
The nonequilibrium charge accumulated on the QD$_1$ is obtained from its lesser Green’s function and is given by~\cite{Wang2001}
\begin{equation}
\delta \rho = \rho - \rho_{\mathrm{eq}}
= \frac{i}{2\pi} \int d\omega
\left[ G^{<}_{11}(\omega) - G^{<}_{11,\mathrm{eq}}(\omega) \right],
\label{eq:B3}
\end{equation}
where $G^{<}_{11}(\omega)$ denotes the electron-electron component of the lesser Green's function of QD$_1$ under nonequilibrium conditions, while $G^{<}_{11,\mathrm{eq}}(\omega)$ denotes its equilibrium counterpart.\\
Now, the charge variation can be decomposed into two physically distinct contributions: an injected part originating from the leads and a screening part associated with the internal potential. To leading order in $V$ and $\theta$, one can write
\begin{equation}
\delta \rho = \left( D_N V + \widetilde{D}_N \theta \right)
- \Pi \, \delta U,
\label{eq:B4}
\end{equation}
where the coefficients
\begin{equation}
D_N = \left( \frac{\partial \rho}{\partial V} \right)_{\mathrm{eq}},
\qquad
\widetilde{D}_N = \left( \frac{\partial \rho}{\partial \theta} \right)_{\mathrm{eq}},
\label{eq:B5}
\end{equation}
are the charge and entropic injectivities, respectively, and
\begin{equation}
\Pi = - \left( \frac{\partial \rho}{\partial U} \right)_{\mathrm{eq}}
\label{eq:B6}
\end{equation}
is the Lindhard response function describing screening. The change in the internal potential is determined by the capacitance relation between the accumulated charge and the electrostatic response~\cite{Hwang2015,Sanchez2013},
\begin{equation}
\delta \rho = C \, \delta U,
\label{eq:B7}
\end{equation}
where $C$ denotes the effective capacitance of QD$_1$. Combining Eqs. \ref{eq:B4} and \ref{eq:B7}, one obtains following equation for the screening potential,
\begin{equation}
\delta U =
\frac{ D_N\,V + \widetilde{D}_N\,\theta}
{C + \Pi},
\label{eq:B8}
\end{equation}
from which the characteristic potentials follow directly as
\begin{equation}
u_N = \frac{D_N}{C + \Pi},
\qquad
z_N = \frac{\widetilde{D}_N}{C + \Pi}.
\label{eq:B9}
\end{equation}
These quantities naturally include both particle and hole contributions due to the presence of superconductivity and can be calculated as,
\begin{equation}
D_N = -\cfrac{e\,\Gamma_N}{2\pi} \int d\omega
\left(-\frac{\partial f(\omega)}{\partial \omega}\right)
\left[
 \left|G^r_{11}(\omega)\right|^2
- \left|G^r_{13}(\omega)\right|^2
\right]_{\mathrm{eq}},
\label{eq:B10}
\end{equation}
\begin{equation}
\widetilde{D}_N = -\cfrac{\Gamma_N}{2\pi} \int d\omega\;
\frac{\omega}{T}
\left(-\frac{\partial f(\omega)}{\partial \omega}\right)
\left[
\left|G^r_{11}(\omega)\right|^2
+ \left|G^r_{13}(\omega)\right|^2
\right]_{\mathrm{eq}},
\label{eq:B11}
\end{equation}
\begin{equation}
\begin{aligned}
\Pi = \frac{1}{2\pi} \int d\omega \, f(\omega)
&\Bigg[
\Gamma_N \Bigg(
\frac{\delta |G^r_{11}(\omega)|^2}{\delta U}
+ \frac{\delta |G^r_{13}(\omega)|^2}{\delta U}
\Bigg)+\frac{\Gamma_S |\omega|}{\sqrt{\omega^2 - \Delta^2}} \,
\Theta(|\omega| - \Delta)
\Bigg(\frac{\delta |G^r_{11}(\omega)|^2}{\delta U}
+ \frac{\delta |G^r_{13}(\omega)|^2}{\delta U}\Bigg)\\
&-\frac{\Gamma_S\Delta}{\sqrt{\omega^2 - \Delta^2}} \,
\Theta(|\omega| - \Delta)\Bigg(
\frac{\delta}{\delta U}
\big\{
G^r_{11}(\omega)\,[G^r_{13}(\omega)]^{*}
\big\}
+ \frac{\delta}{\delta U}
\big\{
G^r_{13}(\omega)\,[G^r_{11}(\omega)]^{*}
\big\}
\Bigg)
\Bigg]_{\mathrm{eq}}.
\label{eq:B12}
\end{aligned}
\end{equation}
These expressions show explicitly how the internal potential responds to external perturbations through a balance between charge injection and screening. In particular, the presence of Andreev processes modifies both injectivities and the Lindhard function, leading to a qualitatively different screening behavior compared to purely normal systems. The charge injectivity $D_{N}$ is governed by the competition between the normal electronic contribution, $\left|G_{11}^{r}\right|^{2}$, and the anomalous particle-hole conversion contribution, $\left|G_{13}^{r}\right|^{2}$, which enter Eq.~\eqref{eq:B10} with opposite signs. Physically, the normal electronic contribution tends to increase the
electronic charge accumulated on QD$_1$, whereas the particle-hole conversion contribution partially compensates this charge through Andreev processes. In contrast, although both contributions enter Eq.~\eqref{eq:B11} with the same sign, the anomalous contribution vanishes identically for the isoelectric case because of particle-hole symmetry. Consequently, under pure thermal driving, the entropic injectivity $\widetilde{D}_N$ is determined entirely by the normal electronic contribution.\\
Following previous works~\cite{Hwang2015,Michalek2015,Szukiewicz2016}, we consider the weak capacitive coupling limit ($C\to0$), where the geometric capacitance is negligible and the response is governed primarily by electronic screening described by the Lindhard function $\Pi$. In this limit, the system remains charge neutral ($\delta \rho = 0$), and the internal potential adjusts to compensate any injected charge. Finally, we emphasize that the present approach is restricted to the weakly nonlinear regime, where the nonequilibrium internal potential is expanded to first order in the applied voltage and thermal biases, thereby capturing the leading effects of nonequilibrium charge redistribution and screening on electrical and thermoelectric transport.
\nocite{*}
 \bibliographystyle{apsrev4-2} 
\bibliography{refs}
\end{document}